\documentclass [MS] {uclathes}

\usepackage[
pdftitle={Debleena Sengupta -- MS Thesis},
pdfauthor={Debleena Sengupta},
pdftex, bookmarksnumbered=true, pagebackref=true]{hyperref}
\hypersetup{colorlinks,
citecolor=[rgb]{0.0,0.3,0.0},
filecolor=[rgb]{0.6,0.0,0.0},
linkcolor=[rgb]{0.0,0.0,0.4},
urlcolor=[rgb]{0.0,0.0,0.4}}

\usepackage{natbib}
\usepackage[]{algorithm2e}
\usepackage{tabularx}

\makeatletter
\def\thebibliography#1{\chapter*{References\markboth
  {REFERENCES}{REFERENCES}}
  \addcontentsline{toc}{chapter}{References}
  \renewcommand\baselinestretch{1}
  \@normalsize
  \list{[\arabic{enumi}]}{\settowidth\labelwidth{[#1]}\leftmargin\labelwidth
    \advance\leftmargin\labelsep
    \usecounter{enumi}}
    \def\newblock{\hskip .11em plus .33em minus -.07em}
    \sloppy
    \sfcode`\.=1000\relax}
\makeatother

\usepackage{graphicx, caption, subcaption}
\graphicspath{ {Images/} }

\usepackage{setspace} 

\usepackage{bm,amsmath,amsthm}

\usepackage{booktabs}
\usepackage{tikz}

\usepackage{url}

\title          {Deep Learning Architectures for Automated Image Segmentation}
\author         {Debleena Sengupta}
\department     {Computer Science}
\degreeyear     {2019}

\chair          {Demetri Terzopoulos}
\member         {Song-Chun Zhu}
\member         {Fabien Scalzo}

\dedication {%
To my mother, Chandana Sengupta, and my father, Dipanjan Sengupta, for their unconditional love and support and for always inspiring me to be the best version of myself.

}

\acknowledgments { I wish to thank my committee members who were
generous with their time and expertise on the subject matter. A
special thanks to Dr.~Demetri Terzopoulos, my committee chairman for
all his help and support through my Master's thesis.

I would like to thank my collaborator Ali Hatamizadeh for his suggestions and guidance through this process. I am thankful for all the opportunities I received through this collaboration.

My sincere thanks to Chris Pagnotta for being extremely encouraging and supportive throughout this research work. 

Finally, I would like to thank my family and friends for reading and editing drafts of this thesis. You are all precious gems in my life.
}

\publication{
\item Hatamizadeh, A., Sengupta, D., Terzopoulos, D., “An End-to-End Trainable Deep Active Contour Layer,” Submitted to the International Conference on Machine Learning., June 2019.
  \item Hatamizadeh, A., Hoogi, A., Lu, W., Sengupta, D., Wilcox, B., Rubin, D., Terzopoulos, D., “Deep Level-Set Active Contours for Lesion Segmentation,” Submitted to the IEEE Transcactions on Medical Imaging (TMI).,June 2019.

}

\abstract { Image segmentation is widely used in a variety of computer
vision tasks, such as object localization and recognition, boundary
detection, and medical imaging. This thesis proposes deep learning
architectures to improve automatic object localization and boundary
delineation for salient object segmentation in natural images and for
2D medical image segmentation.

First, we propose and evaluate a novel dilated dense encoder-decoder
architecture with a custom dilated spatial pyramid pooling block to
accurately localize and delineate boundaries for salient object
segmentation. The dilation offers better spatial understanding and the
dense connectivity preserves features learned at shallower levels of
the network for better localization. Tested on three publicly
available datasets, our architecture outperforms the state-of-the-art
for one and is very competitive on the other two.

Second, we propose and evaluate a custom 2D dilated dense UNet
architecture for accurate lesion localization and segmentation in
medical images. This architecture can be utilized as a stand alone
segmentation framework or used as a rich feature extracting backbone
to aid other models in medical image segmentation. Our architecture
outperforms all baseline models for accurate lesion localization and
segmentation on a new dataset. We furthermore explore the main
considerations that should be taken into account for 3D medical image
segmentation, among them preprocessing techniques and specialized loss
functions. }

\begin {document}
\makeintropages

%

\chapter{Introduction}


\begin{figure}
\centering
\begin{tikzpicture}
\node (A) {\includegraphics[width=\linewidth]{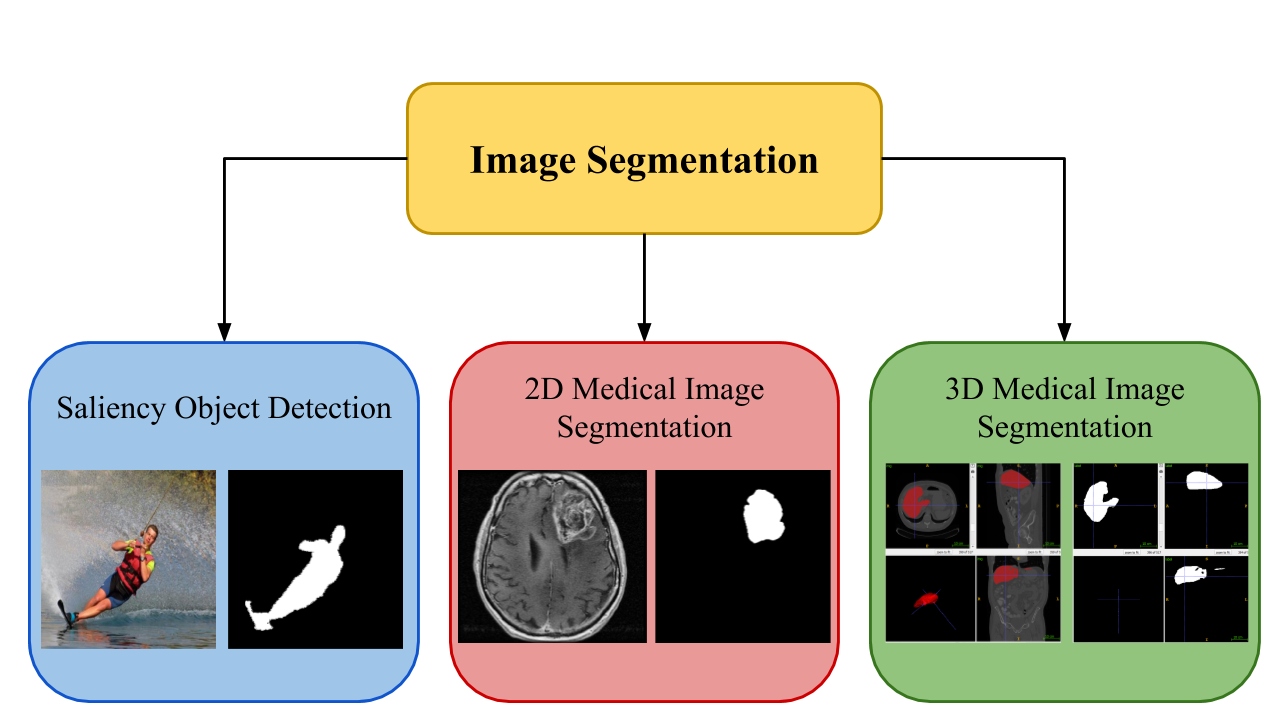}};
\path (A.south west) -- (A.south east) node[pos=.18, below] {(a)} node[pos=.52, below] {(b)} node[pos=.83, below] {(c)};
\end{tikzpicture}
\caption[Types of image segmentation discussed in this work.]{(a) Example of natural image (left) and its corresponding salient object map (right). (b) Example of 2D brain MRI (left) and the corresponding lesion segmentation map (right). (c)Example of 3D liver CT from top, side and front perspectives (left) and corresponding liver segmentation maps for top, side and front perspectives (right). Liver is highlighted in red on the original CT (left).} 
  \label{fig:img_seg}
\end{figure}

Image segmentation is a prominent concept of computer vision. It is
the process of partitioning the pixels of an image into ``segments''
associated with different classes. The main goal of segmentation is
often to simplify the representation of an image such that it is
easier to interpret, analyze, and understand. Image segmentation has
been used in a variety of computer vision tasks, such as object
localization, boundary detection, medical imaging, and recognition. In
essence, these tasks are performed by assigning each pixel in an image
to a certain label based on similar attributes, such as texture,
color, intensity, or distance metrics. The result of image
segmentation is a set of segments that collectively cover the entirety
of an image. The focus of this thesis is to develop novel deep learning methods
for the segmentation of natural images and of 2D and 3D medical
images. Deep learning is a form of machine learning that employs
artificial neural network architectures with many hidden layers
\citep{hinton2012imagenet}.

Specifically, we consider two types of segmentation. The first is
salient object segmentation. On natural images, it involves the
detection of the most prominent, noticeable, or important object
within the image. In image understanding, a saliency map classifies
each pixel in the image as part of the salient object or as part of
the background. Figure~\ref{fig:img_seg}a illustrates an example of an
image and its associated saliency map. Salient object segmentation is
especially important because it focuses attention on the most
important aspects of an image. An example use case is smart image
cropping and resizing, where the image is automatically cropped
without losing the most salient imaged objects. Another use case is in
UI/UX design \citep{salient}, where salient object segmentation is
used to understand what parts of a user interface are most useful,
thus assisting in the development of even better user interfaces for a
smoother user experience. By helping automate the process of
determining the most important parts of images, salient object
segmentation supports a number of other product pipelines.

The second type of segmentation that we consider is medical image
segmentation. Accurate medical image segmentation is often the first
step in a diagnostic analysis of the patient and, therefore, a key
step in treatment planning \citep{aggarwal2011role}. An important
topic in medical image segmentation is the automatic delineation of
anatomical structures in 2D or 3D medical images. In our work, we use
binary medical image segmentation to detect lesions in 2D MR scans and
CT scans of the brain and lung, and swollen lymph nodes in 3D CT
scans. The goal, whether for 2D or 3D images, is to create a model
that can accurately segment a lesion, organ, or cancerous region in a
CT or MR scan. Figure~\ref{fig:img_seg}b and c shows examples of
lesion segmentation in 2D and 3D images. Deep learning approaches to
medical image segmentation raise many challenges not faced in salient
object detection on natural images. Some examples are the lack of
large training datasets, specialized preprocessing techniques on
different medical imaging modalities, memory constraints with 3D
medical images, and the inability to use priors on the shapes of
lesions because they are unique to each patient.

Traditionally, active contour models were broadly applied to the task
of image segmentation
\citep{kass1988snakes,zhu1996region,osher2001level}. These models rely
on the content of the image and minimize an energy functional
associated with deformable contours that delineate the segmentation
process. In recent years, with ever-increasing computational power
from GPUs and the availability of more plentiful training data,
deep convolutional neural networks have achieved state-of-the-art
performance on benchmark datasets in various image recognition tasks
\citep{krizhevsky2012imagenet,simonyan2014very,he2016deep}. Image
segmentation has also benefited from various fully convolutional
architectures \citep{badrinarayanan2015segnet}, notably
encoder-decoder architectures such as UNet \citep{ronneberger2015u}.

\section{Contributions}

Although deep-learning assisted models for image segmentation have
achieved acceptable results in many domains, these models are yet
incapable of producing segmentation outputs with precise object
boundaries. This thesis aims to advance the current paradigm by
proposing and evaluating novel architectures for the improved
delineation of object boundaries in images.

We first tackle the problem of salient object detection and
segmentation in natural images and propose an effective architecture
for localizing objects of interest and delineating their boundaries.
We then investigate the problem of cancerous lesion segmentation and
address the aforementioned issues in various 2D medical imaging
datasets that contain lesions of different sizes and shapes. Finally,
we explore the preprocessing considerations that should be taken into
account when transitioning from 2D segmentation models to 3D. We
evaluate these considerations by preprocessing an abdominal lymph node
dataset. We then train and fine-tune a 3D VNet model using the
preprocessed dataset.

In addition to the suitability of the proposed architectures as
standalone models, they can also be leveraged in an integrated
framework with energy-based segmentation components to decrease the
dependence on external inputs as well as improve robustness. In
particular, the concepts and 2D architectures developed in this thesis (Chapters~3
and 4) were used as a backbone in integrated frameworks for International Conference on Machine Learning (ICML) 2019, Conference on Computer Vision and Pattern Recognition (CVPR) 2019 \citep{hatamizadehdlac2019},International Conference on Computer Vision (ICCV) 2019 \citep{hatamizadehdcac2019} and Machine Learning in Medical Imaging (MLMI) workshop for MICCAI (accepted) \citep{hatamizadeh2019deep}. 

\section{Overview}

The remainder of this thesis is organized as follows:
\begin{itemize}
\item Chapter~2 reviews related work. We first discuss image
segmentation techniques that were popular prior to deep learning
methods. The transition to deep learning segmentation methods is then
highlighted. Finally, three deep learning architectures that inspired
our work are discussed in detail; namely, UNet, DenseNet, and VNet.

\item Chapter~3 introduces a novel deep neural network architecture
for the task of salient object segmentation. Its features include, an
encoder-decoder architecture with dense blocks and a novel dilated
spatial pyramid pooling unit for better spatial image understanding.

\item Chapter~4 explores a more specific use case for image
segmentation---medical image segmentation. We introduce a novel deep
learning architecture for 2D medical image segmentation on a new
medical image dataset from Stanford University. The highlights of this
architecture are an encoder-decoder network with dense blocks that
performs competitively with the state-of-the-art architectures, yet
with very limited training data.

\item Chapter~5 discusses important considerations required to
transition from a 2D segmentation model to a 3D segmentation model.
The importance of preprocessing techniques in the 3D case is
highlighted. A 3D medical dataset is preprocessed using these
techniques and a 3D VNet is trained and fine-tuned with the processed
data.

\item Chapter~6 presents our conclusions and proposes future work.

\end{itemize}

\chapter{Related Work}

This chapter first reviews segmentation techniques that have been
explored prior to the rise of deep learning for segmentation. Then, we
review three noteworthy deep learning architectures that have inspired
our work, namely UNet, DenseNet, and VNet.

\section{Salient Object Segmentation}

Salient object segmentation is the task of detecting the most
prominent object in an image and classifying each image pixel as
either part of that object or part of the background (a binary
classification task). Prior to deep learning methods, there were five
main categories of salient image segmentation methods, namely
region-based methods, classification methods, clustering methods, and
hybrid methods, as discussed by \citet{norouzi2014medical}, as well as
active contour models. For each method, we will describe the
algorithm, followed by its advantages and disadvantages.

There are two main region-based methods, thresholding and region
growing. Thresholding \citep{bogue2005machine} is a simple
segmentation approach that divides the image into different segments
based on different thresholds of pixel intensity. In other words, this
method assumes that, given an image, all pixels belonging to the same
object have similar intensities, and that the overall intensities of
different objects differ. Global thresholding assumes that the pixel
distribution of the image is bimodal. In other words, there is a clear
difference in intensity between background and foreground pixels.
However, this naive, global method performs very poorly if the image
has significant noise and low variability in inter-region pixel
intensity. More sophisticated global thresholding techniques include
Otsu's Thresholding \citep{qu2010research}. This method assumes a
two-pixel class image in which the threshold value aims to minimize
the intra-class variance. Local thresholding helps this  issue by
dividing images into subimages, determining a threshold value for each
subimage, and combining them to obtain the final segmentation. Local
thresholding is achieved using different statistical techniques,
including calculating the mean, standard deviation, or maximum/minimum
of a subregion. An even more sophisticated method of thresholding is
region growing. This method requires an initial seed that grows due to
similar properties for surrounding pixels, representing an object, as
done by \citet{leung1998contour}. The obvious drawback to this
approach is that it is heavily dependent on the selection of the seeds
by a human user. Human interaction often results in a high chance of
error and differing results from different users.

K-nearest neighbors and maximum likelihood are simple classification
methods that were prominent prior to deep learning. For each image,
pixels are assigned to different object classes based on
classification learned at training time. The advantage of maximum
likelihood classification is that it is easy to train. Its disadvantage is
that it often requires large training sets. Clustering methods attempt
to mitigate the issues faced with classification since they do not
require a training set. They use statistical analysis to understand
the distribution and representation of the data and, therefore, are
considered unsupervised learning approaches. Some examples of
clustering include, k-means, fuzzy C-mean, and expectation
maximization \citep{med_lit_review}. The main disadvantage of
clustering methods is that they are sensitive to noise since they do
not take into account the spatial information of the data, but only
cluster locally.

Hybrid methods are techniques that use both boundary (thresholding)
and regional (classification/clustering) information to perform
segmentation. Since this method combines the strengths of
region-based, classification, and clustering methods, it has proven to
give competitive results among the non-deep-learning techniques. One
of the most popular methods is the graph-cut method. This method
requires a user to select initial seed pixels belonging to the object
and to the background. Then, it uses the minimum cut algorithm on the
graph generated from the image pixels, as discussed by
\citet{boykov2001interactive}. The drawback is the need for a user to
select the initial seeds. This is error prone for low contrast images
and would be more advantageous if it could be automated.

Active contour models (ACMs), introduced by \citet{kass1988snakes}
have been very popular for segmentation, as in
\citep{vese2002multiphase}, and for object boundary detection, as in
\citep{mishra2011decoupled}. Active contour models utilize
formulations in which an initial contour evolves towards the object's
boundary via the minimization of an energy functional. Level set ACMs
can handle large variations in shape, image noise, heterogeneity, and
discontinuous object boundaries, as shown by \citet{5754584}. This
method is extremely effective if the initialization of the contours
and weighted parameters is done properly. Researchers have proposed
automatic initialization methods, such as using shape priors or
clustering methods. Although they eliminate a manual initialization
step, techniques such as shape priors require representative training
data.

For the task of salient object detection, we work with natural images,
which offers certain advantages when compared to working with medical
images. Much larger datasets of natural images are available and the
use of priors is possible, since the shapes of basic objects such as
animals, furniture, or cars, are known. Prior models were used by
\citet{han2015background} who proposed a framework for salient object
detection by first modeling the background and then separating salient
objects from the background.

\section{2D and 3D Medical Image Segmentation}

Modern technology has enabled the production of high quality medical
images that aid doctors in illness diagnosis. There are now several
common image modalities such as MR, CT, X-ray, and others. Medical
image segmentation is often the first step in a diagnosis analysis
plan. \citet{aggarwal2011role} explain that accurate segmentation is a
key step in treatment planning. Such images are often segmented into
organs or lesions within organs. Before the advent of computer vision
in medical image analysis, segmentations were often done manually,
with medical technicians manually tracing the boundaries of different
organs and lesions. This process is not only tedious for technicians
but also error prone, since tracing accurate boundaries by hand can be
rather tricky in some cases.

In the past few decades, many effective algorithms have been developed
to aid the segmentation process \citep{med_lit_review}. The methods
reviewed in the previous section have been applied to medical image
segmentation. However, in the medical domain, copious quantities of
labeled training data are rare, therefore clustering methods like
k-means, which was first applied to medical images by
\citet{korn1998fast}, often do not prove to be robust. Furthermore,
obtaining priors for active contour models is sometimes possible but
is oftentimes difficult in the medical imaging domain. In lesion
segmentation, for example, not only is data limited, but every lesion
is unique to a given patient; therefore, it is difficult to obtain a
prior model for lesions. In order to use the techniques mentioned in
the previous section, researchers would heavily depend on data
augmentation, something that is still common with deep learning
models.

The rise of deep learning techniques in computer vision has been
momentous in the recent past. The first to propose a fully
convolutional deep neural network architecture for semantic
segmentation was \citet{long2015fully}. Deep learning is now being
heavily used in medical image segmentation due to its ability to learn
and extract features on its own, without the need for hand-crafting
features or priors. Convolutional and pooling layers are used to learn
the semantic and appearance information and produce pixel-wise
prediction maps. This work enabled predictions on input images of
arbitrary sizes. However, the predicted object boundaries are often
blurred due to the loss of resolution due to pooling. To mitigate the
issue of reduced resolution, \citet{badrinarayanan2015segnet} proposed
an encoder-decoder architecture that maps the learned low-resolution
encoder feature maps to the original input resolution. Adding onto
this encoder-decoder approach, \citet{yu2015multi} proposed an
architecture that uses dilated convolutions to increase the receptive
field in an efficient manner while aggregating multi-scale semantic
information. \citet{ronneberger2015u} proposed adding skip connections
at different resolutions. In recent years, many 2D and 3D deep
learning architectures for segmentation have demonstrated promising
results \citep{hatamizadeh2018automatic, hatamizadeh2019boundary, hatamizadeh2019kidney, hatamizadeh2019deepvessel}.


\section{Deep Learning Architectures}

In this section, the three architectures that inspired the deep
learning models developed in Chapter~3, 4 and 5 are discussed in
greater detail.

\subsection{UNet}

The UNet architecture introduced by \citet{ronneberger2015u} was one
of the first convolutional networks designed specifically for
biomedical image analysis. This network aimed to tackle two issues
that are specific to the domain in medical image segmentation. The
first is the lack of large datasets in this domain. The goal of this
architecture is to produce competitive segmentation results given a
relatively small quantity of training data. Traditional feed-forward
convolutional neural networks with fully connected layers at the end
have a large number of parameters to learn, hence require large
datasets. These models have the luxury of learning little bits of
information over a vast number of examples. In the case of medical
image segmentation, the model needs to maximize the information
learned from each example. Encoder-decoder architectures such as UNet
have proven to be more effective even with small datasets, because the
fully-connected layer is replaced with a series of up convolutions on
the decoder side, which still has learnable parameters, but much fewer
than a fully-connected layer. The second issue the UNet architecture
tackles is to accurately capture context and localize lesions at
different scales and resolutions.

\begin{figure}
  \includegraphics[width=\linewidth]{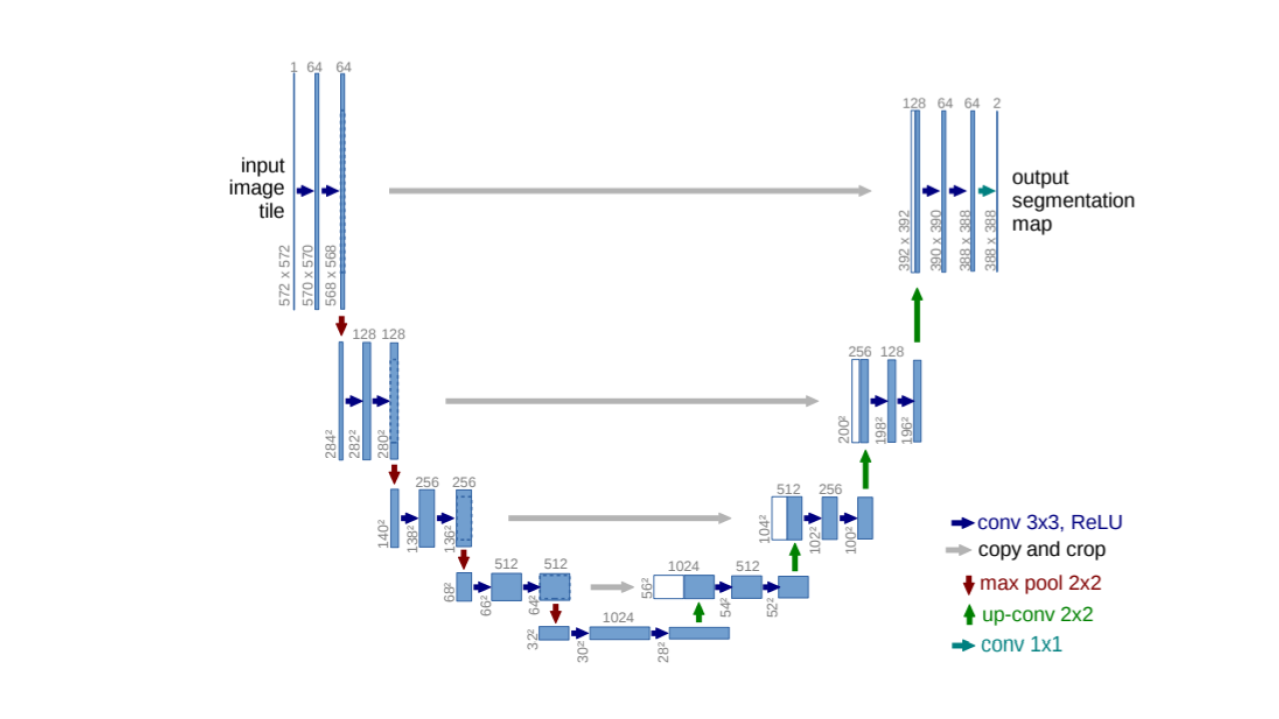}
  \caption{Original UNet architecture from \cite{ronneberger2015u}}
  \label{fig:og_unet}
\end{figure}

\paragraph{Architecture Details:}
The UNet architecture consists of two portions, as shown in
Figure~\ref{fig:og_unet}. The left side is a contracting path, in
which successive convolutions are used to increase resolutions and
learn features. The right side is an expanding path of a series of
upsampling operations. Low resolution features from the contracting
path are combined with upsampled outputs from the expanding path. This
is done via skip connections and is significant in helping gain
spatial context that may have been lost while going through successive
convolutions on the contracting path. The upsampling path utilizes a
large number of feature channels, which enables the effective
propagation of context information to lower resolution layers on the
contracting path, allowing for more precise localization. This
motivated the authors to make the expanding path symmetric to the
contracting path, forming a U-shaped architecture.

The authors also mention another noteworthy issue faced for medical
image segmentation, namely the problem of objects of the same class
touching each other with fused boundaries. To alleviated this issue,
they propose using a weighted loss by separating background labels
between touching segments to contribute to a large weight in the loss
function.

\paragraph{Training:}
The UNet model is trained with input images and their corresponding
segmentation maps via stochastic gradient descent. The authors utilize
a pixel-wise softmax function over the final segmentation and combine
it with a cross-entropy loss function. Weight initialization is
important to help control regions with excessive activation that over
contribute to learning the correct weight parameters and ignoring
other regions. \citet{ronneberger2015u} recommend initializing the
weights for each feature maps so that the network has unit variance.
To achieve this for the UNet architecture, they recommend drawing
the initial weights from a Gaussian distribution. The authors also
discuss the importance of data augmentation since medical image
datasets are often small. Techniques such as shifting, rotating, and
adding noise are most popular for medical images.

The success of the UNet architecture makes it very appealing to
explore and build upon for different medical segmentation tasks.
\citet{ronneberger2015u} demonstrated the success of this model on a
cell segmentation task. We will take this architecture as a baseline
for our work on lesion segmentation in the brain and lung in
Chapter~4.

\subsection{DenseNet}

The DenseNet architecture was introduced by \citet{DenseNet}. Although
this architecture was not designed specifically for application to
medical image segmentation, the ideas can be effectively applied to
the medical imaging domain.
\begin{figure}
  \includegraphics[width=\linewidth]{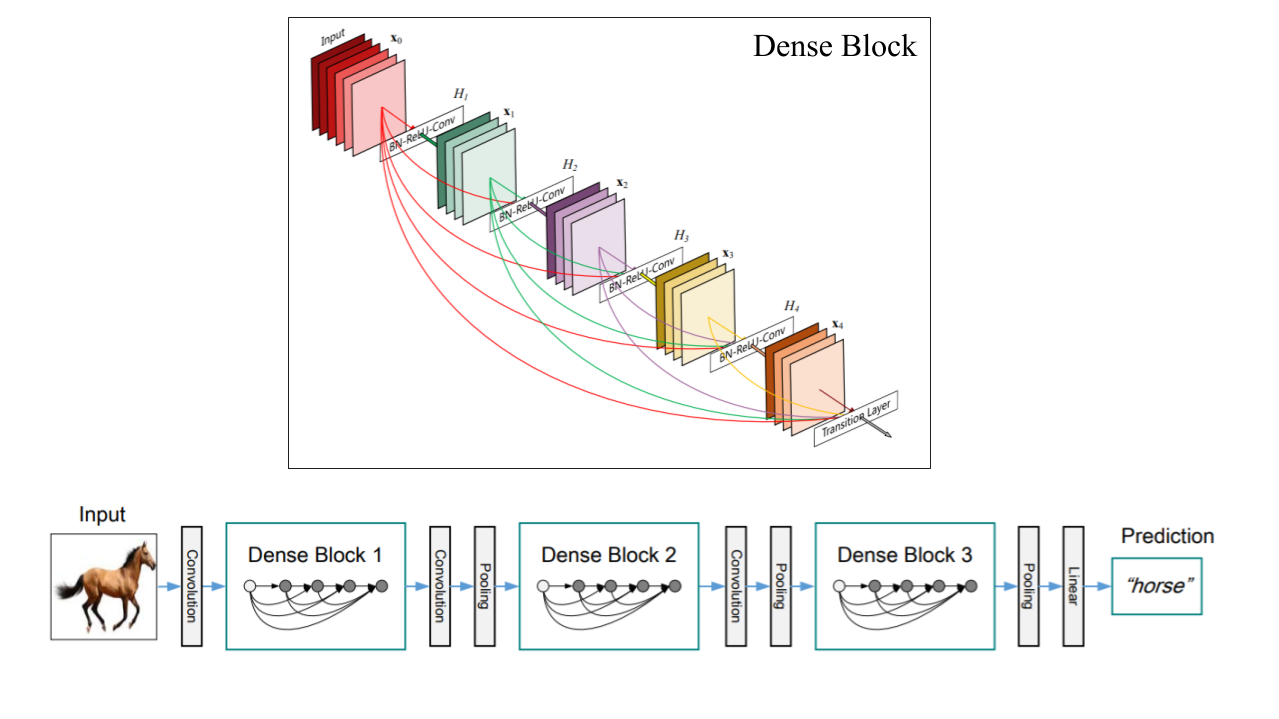}
  \caption[Original DenseNet architecture from \cite{DenseNet}.]{Original DenseNet architecture from \cite{DenseNet}. The top figure depcits an example dense block. The bottom figure depicts a full DenseNet architecture.}
  \label{fig:og_dense}
\end{figure}

\paragraph{Architecture Details:}
DenseNet, depicted in Figure~\ref{fig:og_dense}, is a deep network
that connects each layer to every other layer. This architecture
extends on the observations that deep convolutional networks are
faster to train and perform better if there are shorter connections
between input and output layers. Therefore, each layer takes the
feature maps generated from all preceding layers and the current
feature maps as input for all successive layers. Interestingly enough,
there are actually fewer parameters to learn in this architecture than
in architectures such as ResNet, proposed by \citet{he2016deep},
because it avoids learning redundant information. Instead, each layer
takes what has already been learned before as input. In addition to
efficient parameter learning, the backpropagation of gradients is much
smoother. Each layer has access to the gradients from the loss
function and the original input signal, leading to an implicit deep
supervision. This also contributes to natural regularization, which is
beneficial for small datasets where overfitting is often an issue.

As mentioned previously, a goal of DenseNet is to improve information
flow for fast and efficient backpropagation. For comparison, consider
how information flows through an architecture such as ResNet. In a
traditional feedforward convolutional network, each transition layer
is described as $x_{l} = Z_{l}(x_{x-1})$, where $Z$ is a composite of
convolution, batch norm, and ReLU operations. ResNet also has a
residual connection to the previous layer. The composite function is
\begin{equation}
x_{l} = Z_{l}(x_{x-1}) + x_{l-1}.
\end{equation}
Although the connection to the previous layer assists in gradient
flow, the summation makes gradient flow somewhat slow. To combat this
issue, DenseNet uses concatenation instead of summation for
information flow:
\begin{equation}
x_{l} = Z_{l}([x_{l-1}; x_{l-2}; \dots; x_{0}]),
\end{equation}
where $[a;b;c]$ denotes concatenation. This idea is referred to as
dense connectivity. DenseNet is a collection of such dense blocks with
intermediate convolution and pooling layers, as shown in
Figure~\ref{fig:og_dense}.

The main points of success for this model is that there is no redundancy in learning parameters and the
concatenation of all prior feature maps makes gradient backpropagation
more efficient to assist in faster learning. Ideas from DenseNet have
inspired the work in Chapters~3 and 4.


\subsection{VNet}

\begin{figure}
  \includegraphics[width=\linewidth]{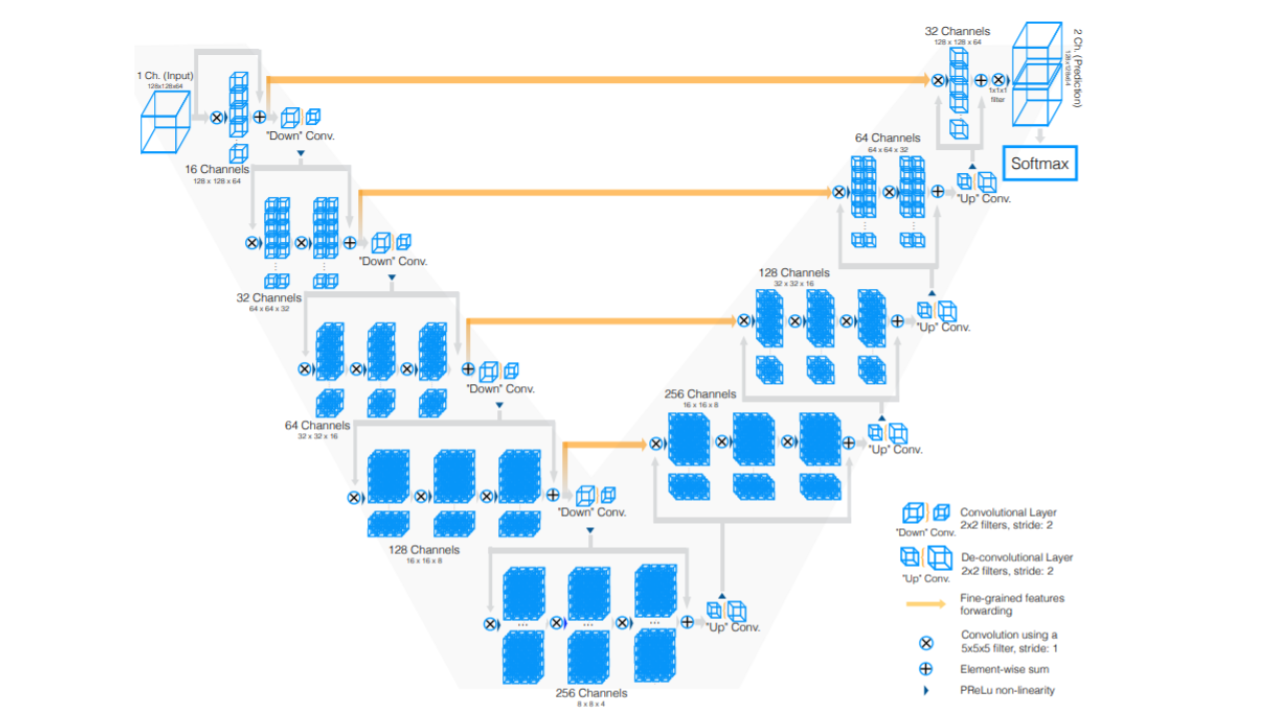}
  \caption{Original VNet architecture from \cite{milletari2016v}.}
  \label{fig:vnet}
\end{figure}

The VNet architecture was introduced by \citet{milletari2016v}. It is
very similar to that of the UNet, as shown in Figure~\ref{fig:vnet}.
The motivation for this architecture was that much of medical image data
is 3D, but most deep learning models at the time seemed to focus on 2D
medical image segmentation. Therefore, the authors developed an
end-to-end 3D image segmentation framework tailored for 3D medical
images.

\paragraph{Architecture Details:}
The VNet architecture consists of a compression path on the left,
which reduces resolution, and an expansion path on the right, which
brings the image back to its original dimensions, as shown in
Figure~\ref{fig:vnet}. On the compression path, convolutions are
performed to extract image features, and at the end of each stage,
reduce the image resolution before continuing to the next stage. On
the decompression path the signal is decompressed and the original
image dimensions are restored.

The main difference between the VNet and the UNet lies in the
operations applied at each stage. In a conventional UNet, as shown in
Figure~\ref{fig:og_unet}, at each stage, the compression path performs
convolutions to extract features at a given resolution and then
reduces the resolution. The VNet does the same, but the input of each
stage is used in the convolutional layers of that stage and it is also
added to the output of the last convolutional layer of that stage.
This is done to enable the learning of a residual function; hence,
residual connections are added in each stage. \citet{milletari2016v}
observed that learning a residual function cuts the convergence times
significantly, because the gradient can flow directly from the output
of each stage to the input via residual connections during
backpropagation.

Another notable difference between the VNet and the UNet is the
technique used to reduce resolutions between stages. In the case of
the UNet, after an input goes through convolutions and a nonlinearity,
it is fed into a max pool layer to reduce resolution. In the case of
the VNet, after the nonlinearity, the input is fed through a
convolution with a $2 \times 2 \times 2$ voxel-wide kernel applied
with a stride of length 2. As a result, the size of the feature maps is halved
before proceeding to the next stage. This strategy is more memory
efficient, which is highly advantageous since memory scarcity is a big
problem for most 3D models.


The expansion portion of the VNet extracts features and enables the
spatial understanding from low resolution feature maps in order to
build a final volumetric segmentation. After each stage of the
expansion path, a deconvolution operation is applied to double the
size of the input, until the original image dimensions are restored.
The final feature maps are passed through a softmax layer to obtain a
probability map predicting whether each voxel belongs to the
background or foreground. Residual functions are also applied on the
expansion path.

The main point of success for this model is that it was the first
effective and efficient end-to-end framework for 3D image
segmentation. It utilized an encoder-decoder scheme with learned
residual functions, resulting in faster convergence than other 3D
networks. It employs convolutions for resolution reduction instead of
using a max pooling layer, which is more memory efficient, a very
important point for 3D architectures. VNet and its variants have
proven to be very promising in a variety of 3D segmentation tasks,
such as multi-organ abdominal segmentation \citep{gibson2018automatic}
and pulmonary lobe segmentation \citep{hatamizadeh2018automatic}. This
architecture will be further explored in Chapter~5.

\chapter{2D Natural Image Segmentation}

The manual delineation of segmentation boundaries is an error-prone
procedure that is cumbersome, time intensive, and subject to user
variability. Deep learning methods automate delineation by having
learned parameters detect boundaries. This chapter discusses the
implementation of a series of deep learning models to improve object
delineation of boundaries in the task of salient object segmentation
in natural images; i.e., images of everyday ``natural'' objects.

Encoder-decoder architectures have proven to be very effective in
tasks such as semantic segmentation; however, they are yet to be
heavily explored for use in salient object detection. To this end, we
propose a custom dense encoder-decoder with depthwise separable
convolution and dilated spatial pyramid pooling. This is a simple and
effective method to assist in object localization and boundary
detection with spatial understanding.

The custom encoder-decoder architecture, as depicted in
Figure~\ref{fig:pipeline_nat}, is tailored to estimate a binary
segmentation map with accurate object boundaries. In this
architecture, the encoder increases the size of the receptive field
while decreasing the spatial resolution by a series of successive
dense blocks. The decoder employs a series of transpose
convolutions and concatenations via skip connections with
high-resolution extracted features from the encoder.

The novelty in our architecture is that we create a light, custom
dilated spatial pyramid pooling (DSPP) block at the end of the
encoder. The output of the encoder is fed into 4 parallel dilation
channels and the results are concatenated before passing it to the
decoder, as shown in Figure~\ref{fig:pipeline_nat}. Since there are
various features in natural images, such as scale and resolution, we
utilized a dilated spatial pyramid pooling block to restore the
spatial information that may have been lost while reducing resolutions
on the encoder side for more accurate boundary detection of the
salient object.

The following sections systematically go through a series of
architectures before arriving at the implementation for the novel
dense encoder-decoder with dilated pyramid pooling. The final
architecture is validated on three prominent publicly available
datasets used for the task of salient object detection, namely MSRA,
ECSSD, and HKU.

\section{Basic Encoder Decoder}


\begin{figure*}
  \includegraphics[width=\textwidth]{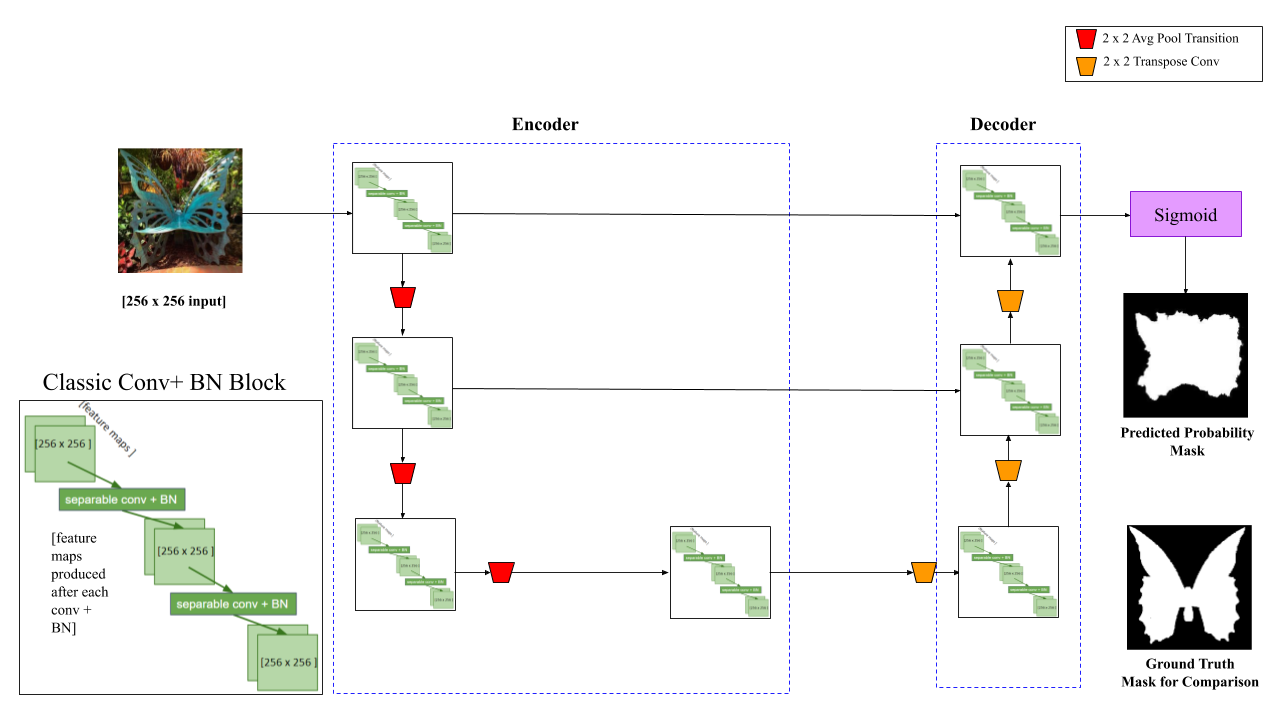}
  \caption[Basic Encoder-Decoder Model.]{Basic Encoder-Decoder Model. The basic encoder-decoder model, a classic convolution + batch norm block and the associated produced feature maps are depicted.}
  \label{fig:baisc_ed}
\end{figure*}

We first explore a basic encoder-decoder architecture, illustrated in
Figure~\ref{fig:baisc_ed}, for the task of salient object detection.

\subsection {Implementation}

In the encoder portion, the number of learned feature maps is
incrementally increased by powers of two, from 64 filters to 1024
filters. The size of the image is decreased by half at each stage of
the encoder. The end of the encoder samples at the lowest resolution
to capture high-level features regarding the object shapes. In the
decoder portion, the number of feature maps is decreased by powers of
two, from 1024 to 64. The image dimensions are restored to the input
dimensions, by doubling at each stage. We pass the last layer through
a sigmoid filter to produce the final binary map.

\paragraph{Preprocessing:}
For this task, the original input image sizes are $300 \times 400$.
The input images are resized to $256 \times 256$. This is done because
256 is the best value to minimize distortion but still make the data
uniform to feed into the network.

\paragraph{Convolution:} 
The basic encoder-decoder architecture utilizes depthwise separable
convolution. Each depthwise separable convolutional layer in the
proposed architecture may be formulated as follows:
\begin{equation}
\label{eq:1st}
c_\mathrm{separable}(X, W, \gamma) = r(b(X * W), \gamma). 
\end{equation}
It consists of an input $X$, a learned kernel $W$, a batch
normalization $b$, and a ReLU unit $r(X) = \max(0, X)$. Depthwise
separable convolution is used because it is a powerful operation that
reduces the computational cost and number of learned parameters while
maintaining similar performance to classic convolutions. Classic
convolutions are factorized into depthwise spatial convolutions over
individual channels and pointwise convolutions, and combines the
results of convolutions over all channels.

\subsection{Results and Analysis}

\begin{table}
\centering
\begin{tabular}{lcccr}
\toprule
dataset &  $F_{\beta}$ & MAE \\
\midrule
ECSSD   &  0.737     & 0.120    \\
MSRA    &  0.805     & 0.093    \\
HKU     &  0.804     & 0.079    \\
\bottomrule
\end{tabular}
\caption[$F_{\beta}$ and MAE scores for basic encoder-decoder
model]{$F_{\beta}$ and MAE scores for basic encoder-decoder model on
the ECSSD, MSRA, and HKU datasets.}
\label{table:encoder-decoder}
\end{table}

As shown in Table~\ref{table:encoder-decoder}, a basic encoder-decoder
structure alone has mediocre performance. Reasons for this could
include that as an image goes deeper into the encoder, the model loses
information learned directly from higher resolutions. As a result, the
model is able to correctly localize the object but is unable to predict accurate image boundaries. This was
especially observed when an image had multiple objects with more
complex boundaries. In the next section, dense blocks are added to the
encoder portion to mitigate this issue.


\section{Dilated Dense Encoder-Decoder}


\begin{figure*}
  \includegraphics[width=\textwidth]{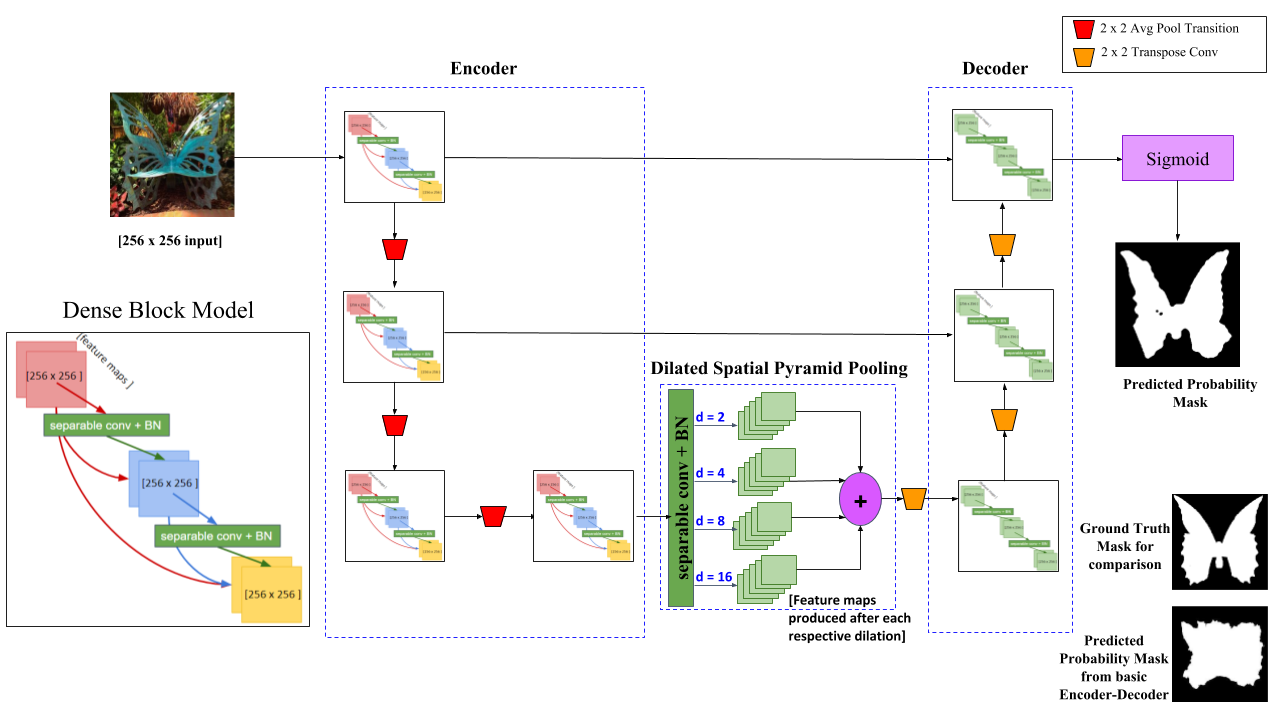}
  \caption[Dilated Dense Encoder-Decoder with Dilated Spacial Pyramid Pooling.]{Dilated Dense Encoder-Decoder with Dilated Spacial Pyramid Pooling. Dilated Dense Encoder-Decoder with Dilated Spacial Pyramid Pooling (DSPP) Block is depicted. Inside the DSPP, the output of the encoder is fed into 4 parallel dilations to increase the receptive fields at different scales, depicted in blue text.}
  \label{fig:pipeline_nat}
\end{figure*}

The main differences between the basic encoder-decoder implementation and the custom encoder-decoder discussed in this section, are as follows:
\begin{itemize}
    \item Implementing dense block units
    \item Implementing a custom dilated spacial pyramid pooling block
\end{itemize}

As shown in Figure~\ref{fig:pipeline_nat}, in addition to using
depthwise separable convolutions, the basic convolutional blocks are
replaced with dense blocks, as employed by \citet{DenseNet}. The
advantage of utilizing dense blocks is that each dense block takes as
input the information from all previous layers. Therefore, at deeper
layers of the network, information learned at shallower layers is not
lost. A custom dilated spatial pyramid pooling block is added at the
end of the encoder for better spatial understanding, as illustrated in
Figure~\ref{fig:pipeline_nat}.

In this dilated dense encoder-decoder model, the encoder portion
increases the receptive field while decreasing the spatial resolution
by a series of successive dense blocks, instead of by convolution and
pooling operations in typical UNet architectures. The bottom of the
dilated dense encoder-decoder consists of convolutions with dilation
to increase the spatial resolution for better image understanding. The
decoder portion employs a series of up-convolutions and concatenations
with high-resolution extracted features from the encoder in order to
gain better localization and boundary detection. The dilated dense
encoder-decoder was designed to allow lower-level features to be
extracted from 2D input images and passed to the higher levels via
dense connectivity for more robust image segmentation.

\subsection {Implementation}

In each dense block, depicted in Figure~\ref{fig:pipeline_nat}, a
composite function of depthwise separable convolution, batch
normalization, and ReLU, is applied to the concatenation of all the
feature maps $[x_{0}, x_{1},\dots, x_{l-1}]$ from layers 0 to $l-1$.
The number of feature maps produced by each dense block is $k + n
\times k$, where each layer in the dense block contributes $k$ feature
maps to the global state of the architecture and each block has $n$
intermediate layers (the $n \times k$ term), plus the $k$ layers that
comprise the transition layer at the end of the dense block (the $k$
term).

We observed that a moderately small growth rate of $k = 12$ sufficed
to learn decent segmentation, allowing us to increase the model's
learning scope with dense blocks. The tradeoff between better learning
via more parameters and still keeping the model relatively
efficient, using a smaller value of $k$, was a consideration during
model tuning for fast convergence. One reason for the success of a
smaller growth rate is that each layer has information about all
subsequent layers and has knowledge of the global state
\citep{DenseNet}. $k$ regulates the amount of new information that is
added to the global state at each layer. This means that any layer in
the network has information regarding the global state of the entire
network. Thus it is unnecessary to replicate this information
layer-to-layer and a small $k$ suffices.

\paragraph{Dilated Spatial Pyramid Pooling:}
Dilated convolutions make use of sparse convolution kernels to
represent functions with large receptive fields and with the advantage
of few training parameters. Dilation is added to the bottom of the
encoder-decoder structure, as shown in Figure~\ref{fig:pipeline_nat}.
Assuming that the last block of the encoder is $x$, and letting
$D(x,d)$ represent the combined batch norm-convolution-reLU function,
with dilation $d$ on input $x$, the dilation may be written as
\begin{equation}
\label{eq:dilate_concat}
Y = [D(x,2); D(x,4);D(x,8); D(x,16)],
\end{equation}
where $[D();D()]$ represents concatenations. The last dense block is
fed into 4 parallel convolutional layers with dilation 2, 4, 8, and
16. Once the blocks go through function $D$, they are concatenated to
gain wider receptive fields and spatial perspective at the end of the
encoder. This is only done at the bottom of the architecture because
it is the section with the least resolution, the ``deepest'' part of
the network. This allows for an expanded spatial context before
continuing into the decoder path.


\subsection{Results and Analysis}

\begin{table}
\centering
\begin{tabular}{lcccr}
\toprule
dataset &  $F_{\beta}$ & MAE \\
\midrule
ECSSD   &  0.825     & 078    \\
MSRA    &  0.857     & 0.061    \\
HKU     &  0.845     & 0.087    \\
\bottomrule
\end{tabular}
\caption[$F_{\beta}$ and MAE score for dilated dense
encoder-decoder]{$F_{\beta}$ and MAE Score for dilated dense
encoder-decoder on the ECSSD, MSRA, and HKU datasets.}
\label{table:dense_dilated_unet}
\end{table}

Table~\ref{table:dense_dilated_unet} shows the results of the dilated
dense encoder-decoder model with dilated spatial pyramid pooling. We
observed that this architecture performed significantly better than
the basic encoder-decoder architecture
(Table~\ref{table:encoder-decoder}).



\begin{table}[ht]
    \centering

   \resizebox{\textwidth}{!}{ \begin{tabular}{lccccccr}
\hline
Model & ECSSD ($F_{\beta}$) & ECSSD (MAE)  & MSRA ($F_{\beta}$) & MSRA (MAE) & HKU ($F_{\beta}$) & HKU (MAE) \\
\hline
\hline
MC  \cite{zhao2015saliency}     & 0.822 & 0.107 & 0.872 & 0.062 & 0.781 & 0.098\\
MDF \cite{DBLP:journals/corr/LiY15} & 0.833  & 0.108 & 0.885 & 0.104 & 0.860 & 0.129\\
ELD \cite{lee2016deep}              & 0.865 & 0.981 & 0.914 & 0.042 & 0.844 & 0.071 \\
ED \cite{ronneberger2015u}          &  0.737 & 0.120 & 0.805 & 0.093 & 0.804 & 0.079 \\
DDED                                &  0.825 & 0.078 & 0.857 & 0.061 & 0.845 & 0.087 \\
DDED + ACL                          & \textbf{0.920}  & \textbf{0.048} & 0.881 & 0.046 & 0.861 & 0.054  \\
SOA \cite{hou2016deeply}            & 0.915 & 0.052 & \textbf{0.927} & \textbf{0.028} & \textbf{0.913} & \textbf{0.039} 
    \end{tabular}}
    \caption[Model Evaluations]{Model Evaluations. $F_{\beta}$ and MAE values for each model on ECSSD, MSRA, and HKU datasets. ED abbreviates encoder-decoder. DDED abbreviates dilated dense encoder-decoder. ACL abbreviates active contour layer.}
    \label{table:datasets-perf}
\end{table}

This network was the backbone CNN used in the full pipeline proposed
in our ICML 2019 submission. The results are presented in
Table~\ref{table:datasets-perf}. DDED indicates results from the
stand-alone dilated dense encoder-decoder, described in this section.
ED indicates the results from the basic encoder-decoder described in the previous section. 

The full pipeline consisted of the DDED backbone CNN and an active contour layer (ACL). The output of the dilated dense encoder-decoder is taken as input into the ACL, which then produces the final segmentation results, seen in (DDED + ACL). As observed in Table \ref{table:datasets-perf}, the DDED + ACL architecture beats the current state-of-the-art for the ECSSD dataset and is competitive with the state-of-the-art for the MSRA and HKU datasets. The ECSSD dataset contained many images with highly complex boundaries. The full DDED + ACL framework accurately delineates object boundaries, resulting in an increased $F_{\beta}$ score for the ECSSD dataset. This demonstrates the strength of this framework, in which the backbone DDED was trained from random initialization but yielded competitive results, in comparison to other models (MC, MDF, ELD) that utilized pre-trained CNN backbones, such as ImageNet \citep{deng2009imagenet}, AlexNet \citep{krizhevsky2012imagenet}, GoogLeNet \citep{szegedy2015going} and OverFeat \citep{sermanet2013overfeat}. 

\begin{figure}
  \centering
  \begin{subfigure}{\textwidth}
  \centering
  \caption{Simple Scene $|$ Center Bias}
    \includegraphics[width=0.7\textwidth]{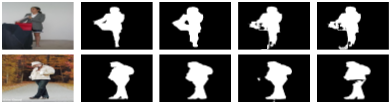}
  \end{subfigure}
  
  \begin{subfigure}{\textwidth}
  \centering
  \caption{Large Object $|$ Complex Boundary}
    \includegraphics[width=0.7\textwidth]{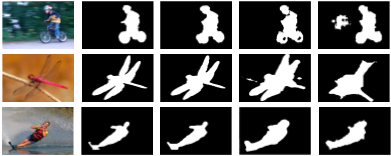}
  \end{subfigure}
  
  \begin{subfigure}{\textwidth}
  \centering
  \caption{Low Contrast $|$ Complex Boundary}
    \includegraphics[width=0.7\textwidth]{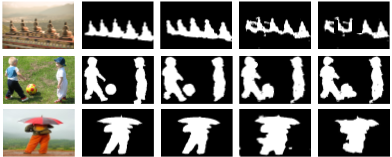}
    
  \end{subfigure}
  
  \begin{subfigure}{\textwidth}
  \centering
  \caption{Large Object $|$ Complex Boundary $|$ Center Bias}
  \begin{tikzpicture}

    \node (A) {\includegraphics[width=0.7\textwidth]{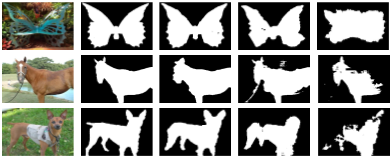}};
\path (A.south west) -- (A.south east) node[pos=.1, below] {(a)} node[pos=.30, below] {(b)} node[pos=.50, below] {(c)} node[pos=.70, below] {(d)} node[pos=.89, below] {(e)};
    \end{tikzpicture}
  \end{subfigure}
  \caption[Examples of Segmentation Outputs.]{Examples of Segmentation Outputs. (a) Original image. (b) Ground truth. (c) DDED + ACL output segmentation. (d) DDED output segmentation. (e) ED output segmentation. }
  \label{fig:compare}
\end{figure}

Figure \ref{fig:compare} illustrates the performance of the ED, DDED, and DDED + ACL frameworks for precise boundary delineation. The images are grouped into categories highlighting different characteristics of the images. The grouping is utilized to indicate the success of the DDED and the DDED + ACL frameworks in a variety of cases. It is observe that all of the images from the dilated dense encoder-decoder (d), are much closer to the ground truth (b) than the results from basic encoder-decoder (e). Therefore (d) proves to be a very robust backbone architecture for this segmentation pipeline, which is required for accurate boundary detection of the ACL. The results in (d) also indicate that the dilated dense encoder-decoder model can also be a successful stand alone model, as almost all images produced are very close to the ground truth (b).


\captionsetup[subfigure]{labelformat=empty}




\section{Loss Function} 

The Dice coefficient is utilized as the loss function, which is
defined as
\begin{equation}
\label{eq:DiceLoss}
\mathrm{Loss} = 1 - \frac{2 \times |X \cap Y|}{|X| + |Y|},
\end{equation}
where $X$ is the prediction matrix, $Y$ is the ground truth matrix,
$|X|$ is the carnality of the set $X$, and $\cap$ denotes
intersection. The Dice coefficient performs better at class-imbalanced
problems by design, by giving more weight to correctly classified
pixels ($2 \times|X \cap Y|$).


\section{Evaluation Metrics}

\begin{figure}
\centering
\begin{tikzpicture}
\node (A) {\includegraphics[width=\linewidth]{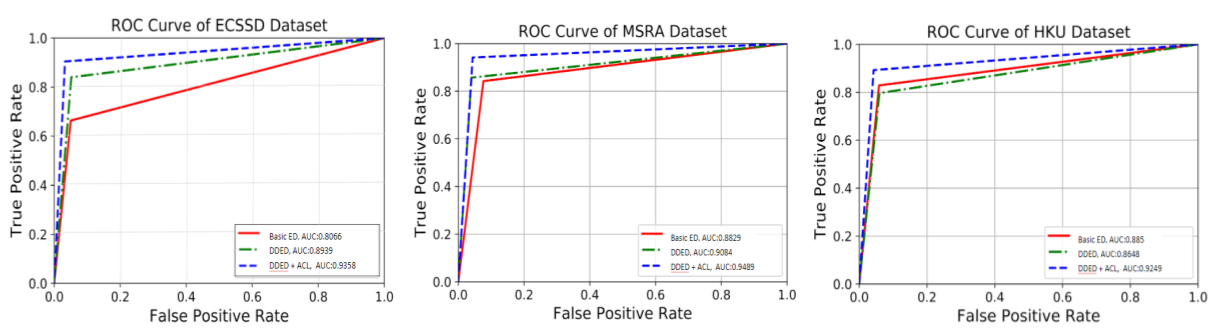}};
\path (A.south west) -- (A.south east) node[pos=.18, below] {(a)} node[pos=.52, below] {(b)} node[pos=.83, below] {(c)};
\end{tikzpicture}
\caption[ROC Performance Curves for Salient Object Detection.]{ROC curves showing the performance of each architecture on ECSSD, MSRA, and HKU dataset.} 
  \label{fig:ROCCurve}
\end{figure}

We utilize three evaluation metrics to validate the model's
performance, namely $F_{\beta}$, ROC curves, and mean absolute error
(MAE). 

\textbf{$F_{\beta}$ Metric}: The $F_{\beta}$ score measures the similarities between labels
and predictions, using precision and recall values, as follows:
\begin{equation}
F_{\beta} = (1+\beta) \times \frac{\mathrm{precision}\times
\mathrm{recall}}{\beta \times \mathrm{precision} + \mathrm{recall}}
\end{equation}

Precision and recall are two metrics that help understand the success of a deep learning model. Precision or recall alone cannot capture the performance of salient object detection. Based on the nature of the datasets being used to validate the model, weights, dictated by the $\beta$ value, will be assigned to precision and recall, accordingly. We use the harmonic weighted average of precision and recall. In the case if salient object detection, there is no need to give more importance to either precision or recall, since all three datasets were rather balanced in terms of class representation. Therefore, we decided to set $\beta = 1$ to
give equal weights to both precision and recall values. The results of
these studies are presented in Table \ref{table:datasets-perf}.


\textbf{ROC Curves}: In addition to the $F_{\beta}$, the ROC
metric is utilized to further evaluate the overall performance boosts that dense blocks and dilated spacial pyramid pooling adds to the salient object detection task. The ROC curves are shown in
Figure \ref{fig:ROCCurve}. A set of ROC curves was created for each of the three
datasets. Each ROC curve consists of the results from testing the
architectures listed in Table~\ref{table:datasets-perf}, namely, a
basic encoder-decoder (ED), the custom dilated dense encoder-decoder (DDED), and the custom dilated dense encoder-decoder + ACL architectures (DDED + ACL). From the
curve trends, it is evident that the dilated dense encoder-decoder + ACL model outperforms the
others due to its high ratio of true positive rate (TPR) to false
positive rate (FPR); i.e., a majority TP cases and few FP cases.
Although ROC curves show the boost in performance gained by using the ACL in our architecture for all three datasets, it is observed that the stand alone dilated dense encoder-decoder architecture performs significantly well in comparison to the basic encoder-decoder model, indicating that it too can perform accurate object boundary delineation. This is observed especially in the ECSSD and MSRA dataset, in Figure \ref{fig:ROCCurve}, since the trends for DDED and DDED + ACL are very close. 

\textbf{Mean Absolute Error}: The mean absolute error metric calculates the amount of difference, or ``error" between the prediction and the ground truth. We utilized the
MAE score for model evaluation, as follows:
\begin{equation}
\label{eq:f_beta}
MAE = \frac{1}{W\times H}\sum_{x=1}^{W}\sum_{y=1}^{H}S(x,y) - G(x,y),
\end{equation}
where $W$ and $H$ are the pixel width and height of the prediction
mask $S$, and $G$ is the ground truth mask, which is normalized to
values $[0,1]$ before the MAE is calculated.



\section{Datasets}

\subsection{Overview}

For the task of salient object detection, three datasets were used,
namely, ECSSD, MSRA, and HKU-IS. Table~\ref{table:dataset_stats} shows
the breakdown of the datasets.

\begin{table*}[b]
\centering
\vskip 0.15in
\begin{center}
\resizebox{\textwidth}{!}{\begin{tabular}{lcccccr}
\hline
dataset & \# Samples Train & \# Samples Valid  & \# Samples Test & Total dataset Size \\
\hline
\hline
ECSSD \citep{yan2013hierarchical}       & 900  & 50 & 100 & 1050\\
MSRA \citep{WangDRFI2017}               & 2700  & 300 & 1447 & 4447\\
HKU-IS \citep{DBLP:journals/corr/LiY16} & 2663  & 337 & 2000 & 5000\\
\end{tabular}}
\end{center}
\caption[MSRA, ECSSD, and HKU dataset breakdowns]{MSRA, ECSSD, and HKU
dataset breakdowns.}
\vskip -0.1in
\label{table:dataset_stats}
\end{table*}

There were some notable differences in the datasets. The MSRA dataset
is mainly a collection of single objects that are centered in the
image. The outlines of most ground truth maps are simple to moderately
complex boundaries. However, there are several images in which the
salient object detection would be difficult, even for a human. These
cases are mainly when objects are partially occluded, making it
difficult to distinguish what object is supposed to be detected. The
ECSSD and HKU datasets had many examples with high-complexity
boundaries. Most images contain multiple salient objects and the
object outlines in the ground truth images were more complex. These
complexities substantially affect the boundary contour evolution. For this
reason, the dilated dense encoder-decoder was developed to capture
varying spatial information and boundaries in the image to give a
solid starting point to be fed into our ACM layer. The test sets are
the same as those reported by \citet{hou2016deeply} for the MSRA and
HKU datasets. Since there was no specified test set for ECSSD, 90\% of
the data as used for training, 5\% for validation, and 10\% for
testing.

\subsection{Data Augmentation and Pretraining}

\begin{figure*}
  \begin{tikzpicture}
    \node (A) {\includegraphics[width=\textwidth]{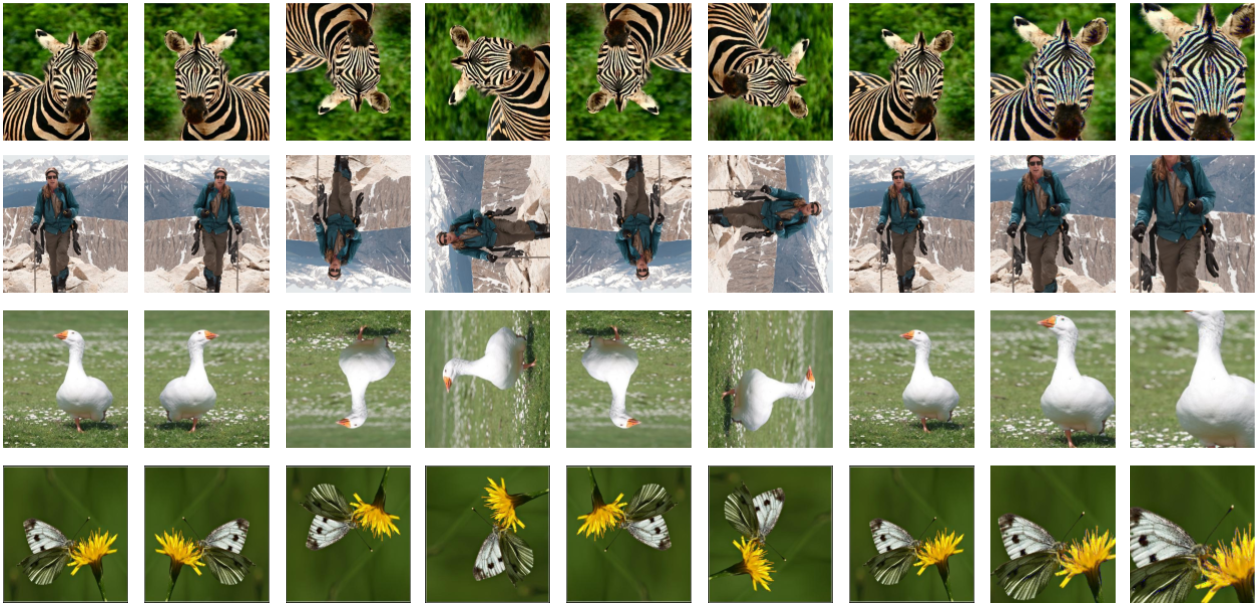}};
\path (A.south west) -- (A.south east) node[pos=.06, below] {(a)} node[pos=.17, below] {(b)} node[pos=.28, below] {(c)} node[pos=.39, below] {(d)} node[pos=.50, below] {(e)} node[pos=.61, below] {(f)} node[pos=0.72, below] {(g)} node[pos=.83, below] {(h)} node[pos=.94, below] {(i)};
    \end{tikzpicture}
    \caption[Examples of Augmentation.]{Examples of Augmentation. (a) is the original image, (b) left-right flip, (c) up-down flip, (d) 90 deg rotation, (e) 180 deg rotation, (f) 270 deg rotation, (g) 1.3 scale zoom, (h) 1.7 scale zoom, (i) 1.9 scale zoom.}
  \label{fig:augment}
\end{figure*}

Table~\ref{table:dataset_stats} shows that the size of each dataset
was considerably small, but these datasets were still utilized because
they were publicly available and highly popular for the task of
salient object detection. But the small size of the dataset did
negatively affect initial attempts at training the models. Therefore
all three datasets were expanded through data augmentation, by
applying to each image the following transformations:
\begin{enumerate}
  \item left-right flip;
  \item up-down flip;
  \item $90^\circ$, $180^\circ$, and $270^\circ$ rotations;
  \item zooming on the image at 3 different scales.
\end{enumerate}
Examples of the augmented dataset are shown in
Figure~\ref{fig:augment}.

With the augmentation, the size of each dataset grew by a factor of 8.
However, training on each individual augmented dataset alone was still
insufficient to fully generalize the model. The data augmentation
helped the ECSSD dataset, so the trained model from this step was used
as a pretrained model for the MSRA dataset. Once the MSRA dataset was
trained, this combined pretrained model was used to train on the HKU
dataset, which was the most challenging due to its more complex
examples.

With this training approach, the results are presented in
Table~\ref{table:datasets-perf}. As can be seen, the DDED and DDED +
ACL results are competitive with the state-of-the-art for these
datasets \citep{hou2016deeply}. \citet{hou2016deeply} use VGGNet
\citep{simonyan2014very} and ResNet-101 \citep{he2016deep} as their
backbones. What is impressive is that our custom dilated dense
encoder-decoder backbone is competitive without utilizing
sophisticated pretrained backbones. This demonstrates that our model is able to train from scratch and produce competitive results with the state-of-the-art and is highly accurate for precise salient object
detection.

\chapter{2D Medical Image Segmentation}

This chapter explores a specific and significant use case of
segmentation, namely 2D medical image segmentation. Medical
images can be 2D or 3D depending on the acquisition equipment. The
main advantages of using a 2D dataset over a 3D dataset is that 2D
images are more memory efficient and 2D models are more lightweight in
terms of the number of learned parameters. Therefore 2D models can
learn at a faster pace for accurate automatic delineation of lesion
boundaries. Since deep learning models require a sizable quantity of
data to generalize due to the large number of learned parameters,
working exclusively with 3D data can be difficult. 3D data can be
sliced into 2D segments to create a larger 2D dataset for models to
learn. The following sections present a number of deep learning models
for the task of 2D lesion segmentation in medical images.

In the past, segmentation of medical images was often manual, cumbersome, time consuming, and often error-prone. Early
computer-assisted segmentation methods required less human
interaction, but still required a user to initialize contours. The
main objective of this chapter is the development of a deep learning
model for fully automatic delineation of lesion boundaries in medical
images, in particular a novel 2D dilated dense UNet architecture for
brain and lung segmentation. Accurate automated segmentation
frameworks can be of great assistance in the early stages of medical
image analysis and the detection of health issues.

\begin{figure}
\centering
\begin{tikzpicture}
\node (A) {\includegraphics[width=\linewidth]{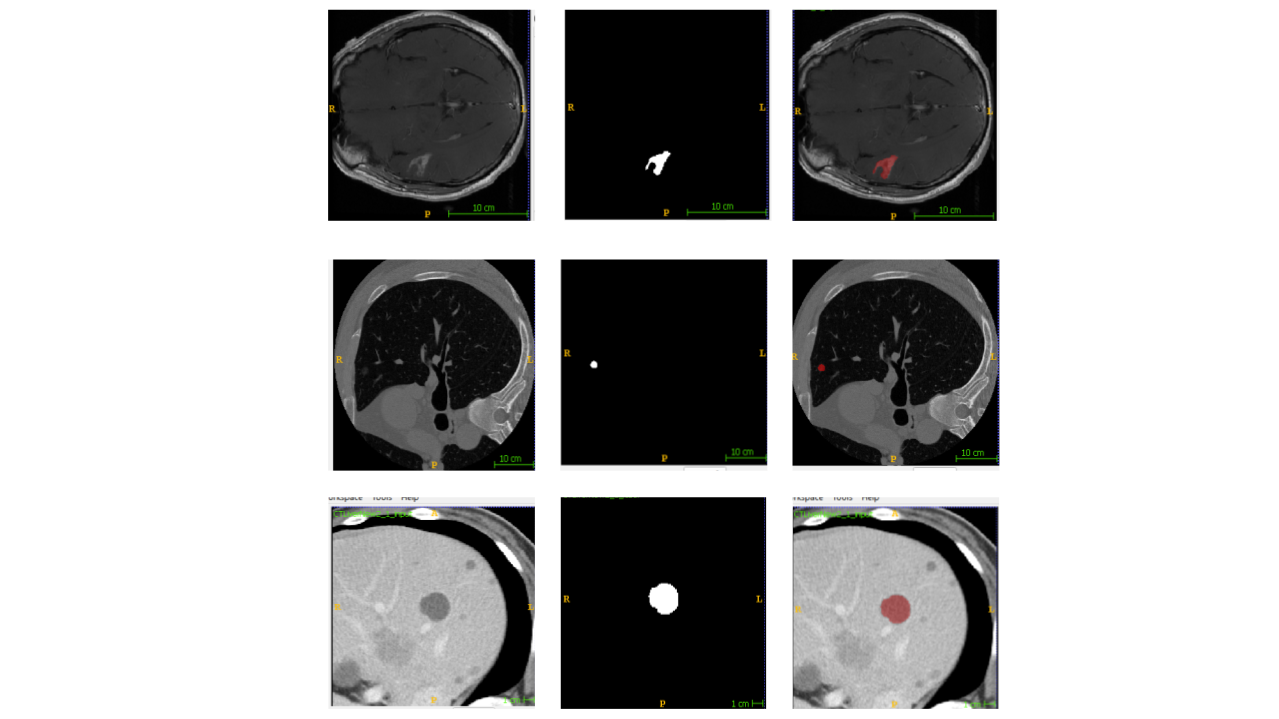}};
\path (A.south west) -- (A.south east) node[pos=.34, below] {(a)} node[pos=.52, below] {(b)} node[pos=.69, below] {(c)};
\end{tikzpicture}
\caption[Examples of Brain, Lung, and Liver Segmentation.]{Examples of brain, lung, and liver (row order) images. (a) original image, (b) ground truth segmentation, (c) overlay of ground truth on original image.}
  \label{fig:2D_seg_pics}
\end{figure}

However, this is a challenging task due to a number of factors, such
as low-contrast images making boundary detection difficult and the
inability to use priors for lesion segmentation, among others.
Figure~\ref{fig:2D_seg_pics} shows example images of brain, lung, and
liver to demonstrate how challenging the segmentation task can be. For
this task, a custom dataset of MR and CT scans is used to detect
lesions in the brain and lung. This dataset was developed in
collaboration with Stanford University. Since this was not a publicly
released dataset, the baseline, for performance comparison, was the results from a basic UNet
model, discussed in the next section.


\section{Baseline UNet}

\begin{figure}
  \includegraphics[width=\linewidth]{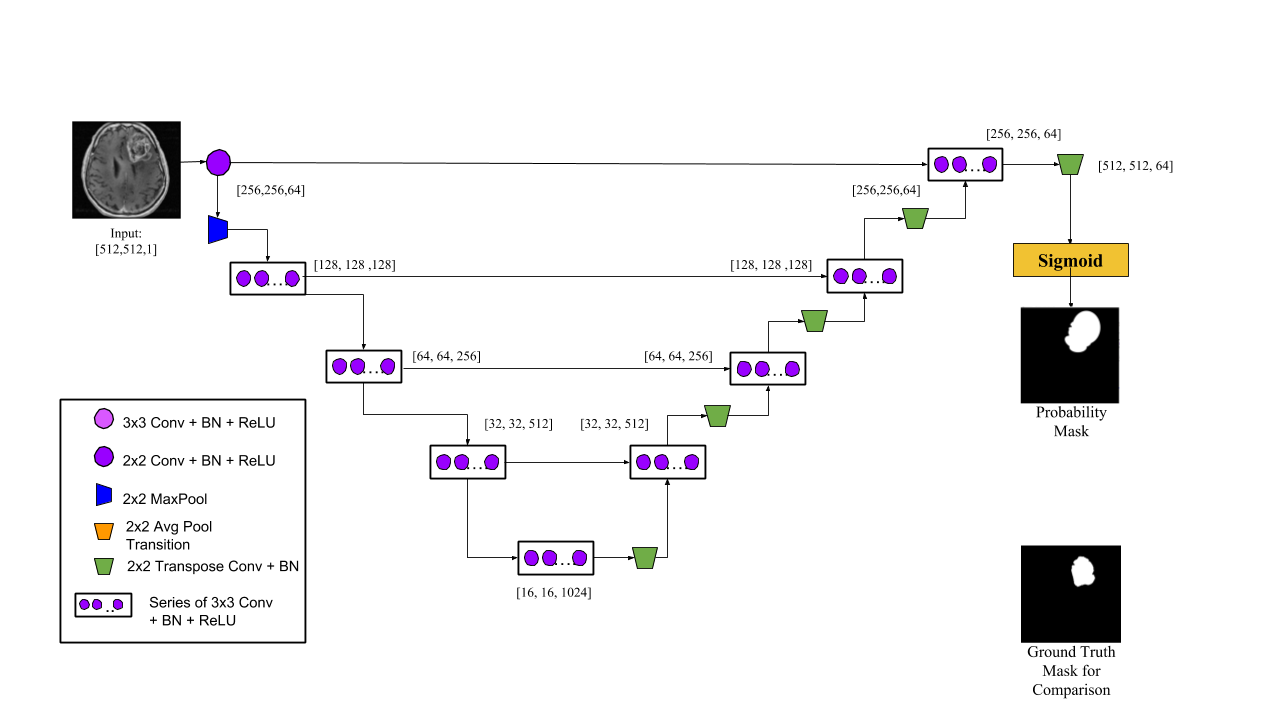}
  \caption[Baseline UNet Model]{The baseline UNet model.}
  \label{fig:unet}
\end{figure}

The UNet architecture was first introduced by \citet{ronneberger2015u}
as a method to perform segmentation for medical images. The advantage
of this architecture over other fully connected models is that it
consists of a contracting path to capture context and a symmetric
expanding path that enables precise localization and automatic
boundary detection with fewer parameters than a feed-forward network.
Therefore this model was successful on small medical image datasets.
The basic UNet architecture is visualized in Figure~\ref{fig:unet}

\begin{table}
\centering
\begin{tabular}{lcccr}
\toprule
Organ & Modality & Model: UNet \\
\midrule
Brain   & MR      & 0.5231      \\
Lung    & CT       & 0.6646      \\
\bottomrule
\end{tabular}
\caption[Dice score for the baseline UNet model on the brain and lung
datasets]{Dice score for the baseline UNet model on the brain and lung
datasets.}
\label{table:Unet}
\end{table}

\subsection{Implementation}

A basic UNet architecture was implemented to provide a baseline
performance benchmark on the Stanford dataset. The encoder portion was
implemented by incrementally increasing the number of feature maps by
powers of two, from 64 filters to 1024 filters at the bottom of the
``U'', at the lowest resolution, to capture intricate details
regarding the lesion shapes. In the decoder portion, we symmetrically
decrease the number of feature maps by powers of 2, from 1024 to 64.
The final image is passed through a sigmoid layer to produce the final
binary segmentation map.

\paragraph{Convolution:}
Each convolutional layer in the proposed architecture consists of a
learned kernel $W$, a batch normalization, and a ReLU unit $r(X) = \max(0, X)$; that is:
\begin{equation}
\label{eq:conv}
c(X, W, \gamma) = r(b((X * W)), \gamma),
\end{equation}
where batch normalization $b(X,\gamma)$ transforms the mean of each
channel to 0 and the variance to a learned per-channel scale parameter
$\gamma$. The ReLU unit introduces non-linearity and assists in
gradient propagation. Each convolutional block consists of a series of
$c(X, W, \gamma)$ layers, as demonstrated in Figure~\ref{fig:unet}.

\subsection{Results and Analysis}

Table~\ref{table:Unet} shows the results of the UNet model. It is
clear that the UNet architecture alone is not enough to perform
accurate segmentation. This could be due to the fact that some lesions
are so small that as the encoder reduces resolution, shape information
is lost, which hinders the ability of the model to pick up detailed
lesion shapes. Therefore, we will improve the model by replacing the
basic convolutional blocks of the UNet with dense blocks.

\section{Dense UNet}

In the next iteration of the model, the convolutional blocks are
replaced with dense blocks, as described by \citet{DenseNet}. The
advantage of utilizing dense blocks is that each dense block is fed
information from all previous layers, as was discussed in
Section~2.3.2. Therefore, information learned from shallower layers is
not lost by the deeper layers. Dense blocks consist of bottleneck
layers. To move from one dense block to the next in the network,
transition layers are utilized. The
implementation of the bottleneck and transition layers are described
in more detail in the next section.

\subsection{Implementation}

\begin{figure}
  \includegraphics[width=\linewidth]{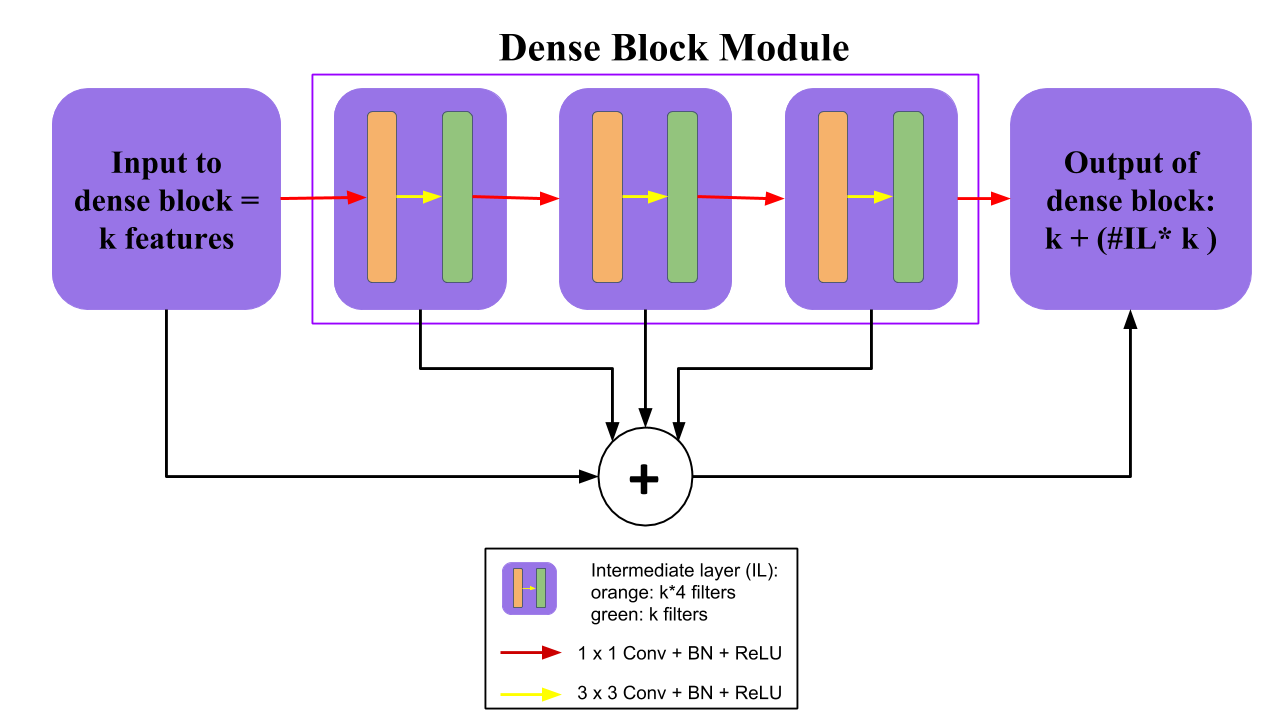}
  \caption[Dense Block Module]{Dense block module. $k$ is the growth
  rate. Before every dense block there are $k$ feature maps generated
  by the transition layer. Within the dense block, there are $n$
  intermediate layers (IL), each contributing $k$ feature maps (green
  block). The total number of feature maps produced is $k + (n \times
  k)$ or, the sum of the input maps and $k$ times the number of
  intermediate layers.}
  \label{fig:denseBlock}
\end{figure}

\paragraph{Dense Blocks:}
The classic convolution blocks in UNets are replaced with a version of
dense blocks. Figure~\ref{fig:denseBlock} illustrates the
implementation of the dense block module. Dense blocks take in all
features learned from previous layers and feed it into subsequent
layers via concatenation. Dense connectivity can be formulated as
follows:
\begin{equation}
\label{eq:dense_conv}
X_{l}  = H_{l}([x_{l0}, x_{l1}, . . . , x_{l-1}]),
\end{equation}
where $H()$ can be considered a composite function of batch
normalization, convolution, and ReLU unit, and $[x_{0}, x_{1}, . . . ,
x_{l−1}]$ represents the concatenation of all feature map from layers
0 to $l-1$. This is a more memory efficient because the model is not
learning redundant features by duplicating feature maps. Direct
connections are implemented between feature maps learned at shallow
levels to deeper levels.

The number of feature maps that are generated by each dense block is
dictated by a parameter called the growth rate. For any dense block
$i$, the number of feature maps is calculated as
\begin{equation}
\label{eq:num_feature_maps}
f_{i} = k + (k \times n),
\end{equation}
where $k$ is the growth rate and $n$ is the number of dense
connections to be made. Figure~\ref{fig:denseBlock} presents a visual
representation of the feature maps in a dense block. Before every
dense block $k$ feature maps are generated by the transition layer.
The growth rate regulates how much new information each layer
contributes to the global state. Within the dense block itself, $n$
connections are made; hence, the total number of filters is the sum of
the two values. It was found that for smaller datasets, smaller values
of $k$ (16--20) suffice to learn nuances of the data without
overfitting, but perform better than a standard UNet.

\paragraph{Bottleneck Layer:}
The dense UNet model has a moderate number of parameters despite
concatenating many residuals together, since each $3\times3$
convolution can be augmented with a bottleneck. A layer of a dense
block with a bottleneck is as follows:
\begin{enumerate}
  \item Batch normalization;
  \item $1\times1$ convolution bottleneck producing growth rate
  $\times 4$ feature maps;
  \item ReLU activation;
  \item Batch normalization;
  \item $3\times3$ convolution producing growth rate feature maps;
  \item ReLU activation.
\end{enumerate}

\paragraph{Transition Layer:}
Transition layers are the layers between dense blocks. They perform
convolution and pooling operations. The transition layers consist of a
batch normalization layer and a $1 \times 1$ convolutional layer
followed by a $2 \times 2$ average pooling layer. The transition layer
is required to reduce the size of the feature maps by half before
moving to the next dense block. This is useful for model compactness.
A transition layer is as follows:
\begin{enumerate}
  \item Batch normalization;
  \item $1\times1$ convolution;
  \item ReLU activation;
  \item Average pooling.
\end{enumerate}

\begin{figure}
  \includegraphics[width=\linewidth]{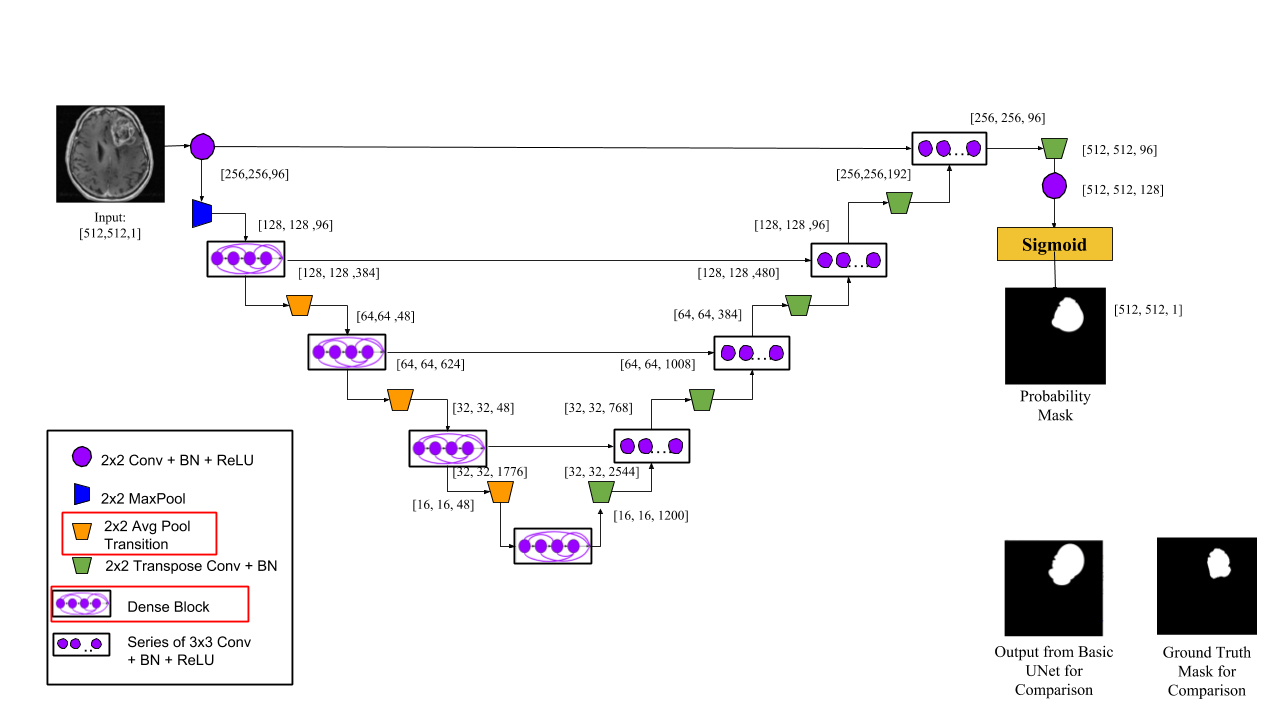}
  \caption[Dense UNet Model]{Dense UNet model. In this design,
  convolutional blocks from Figure \ref{fig:unet} are replaced with dense
  block on the encoder side. Transition layers are added for model
  compactness. The new modifications to this model are highlighted in
  red in the key.}
  \label{fig:densunet}
\end{figure}

Figure~\ref{fig:densunet} illustrates the full dense UNet
architecture.

\begin{table}
\centering
\begin{tabular}{lcccr}
\toprule
Organ & Modality & Model: Dense UNet \\
\midrule
Brain   & MR      & 0.5839     \\
Lung    & CT       & 0.6723     \\
\bottomrule
\end{tabular}
\caption[Dice score for the dense UNet model on the brain and lung
datasets]{Dice score for the dense UNet model on the brain and lung
datasets.}
\label{table:DenseUnet}
\end{table}

\subsection{Results and Analysis}

Table~\ref{table:DenseUnet} shows the results of the dense UNet model.
There are many advantages of using the dense block modules. Recent
work shows that with deeper convolutional networks, prediction
accuracy is increased by creating shorter connections to layers close
to both the input and the output, so that information is not lost as
the network reaches deeper layers. In the standard UNet, let us assume
each convolutional block has $L$ layers. This means there are only $L$
connections (one between each layer; i.e., the output of one layer is
the input of next, and the next layer does not have information about
layers prior to its immediate neighbor). In the case of dense blocks,
$L\times (L+1)/2$ direct connections are being fed into the next block
(i.e., direct, shorter connections to the input and output). This is
advantageous for several reasons. Because of these direct connections,
the vanishing gradient problem is alleviated, and there is stronger
feature propagation so that deeper layers do not lose information
learned early on in the network as the resolution decreases. This also
helps with memory. With careful concatenation, deeper layers have
access to feature maps of shallow layers with only one copy of these
feature maps in memory, instead of multiple copies.

\section{Dilated Dense UNet}

\begin{figure}
  \includegraphics[width=\linewidth]{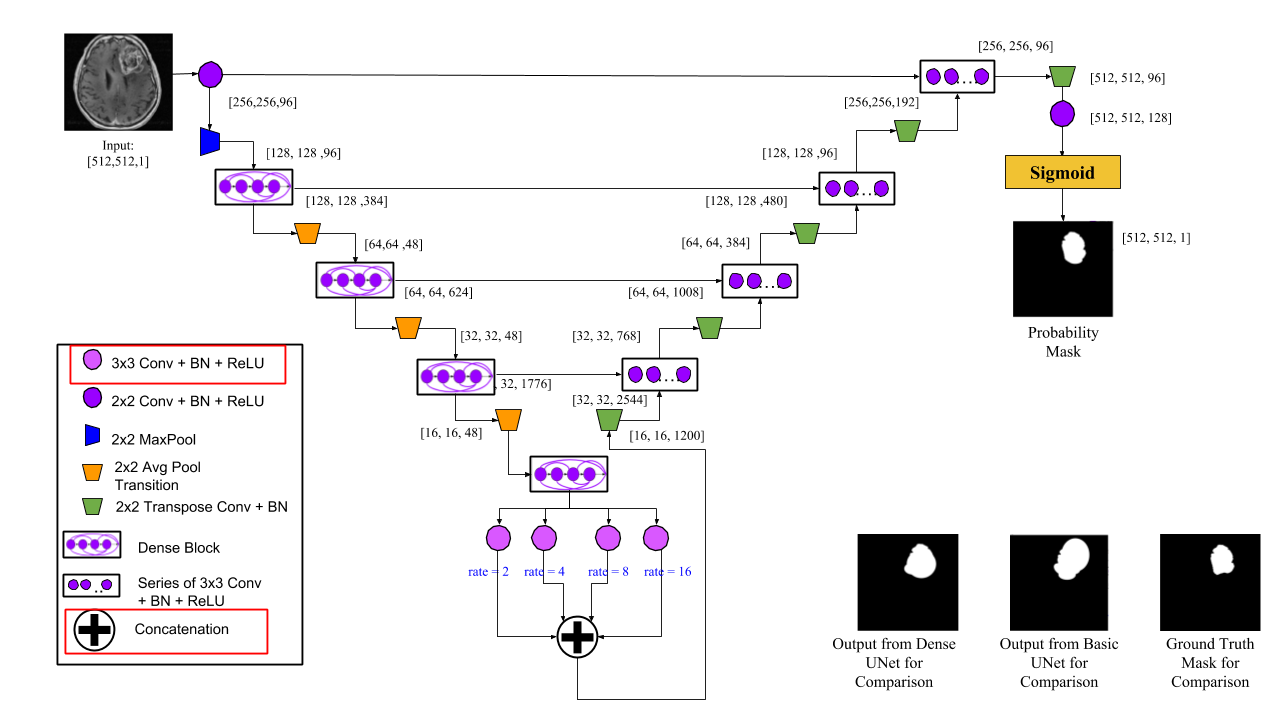}
  \caption[Dilated Dense UNet Model]{Dilated Dense UNet model. In this
  design, dilation is added to the bottom of the network. The new
  modifications to this model are highlighted in red in the key.}
  \label{fig:dilateddensunet}
\end{figure}

Next, we implement a 2D dilated dense UNet for predicting a binary
segmentation of a medical image. This architecture is described in
Figure~\ref{fig:dilateddensunet}. First, the traditional convolution
blocks of a UNet are replaced with dense blocks. Second, an
up-convolution method with learnable parameters is utilized. Third,
dilation is added at the bottom of the architecture for better spatial
understanding.

As can be seen in Figure~\ref{fig:dilateddensunet}, the structure
remains similar to that of the UNet; however, there are key
differences that enable this model to outperform the basic UNet for
the medical image segmentation task. Unlike feed-forward convolutional
neural networks, in which each layer only receives the feature maps
from the previous layer, for maximal information gain per convolution,
every layer of the dilated dense UNet structure takes as input all the
feature maps learned from all the previous layers via dense blocks.
This results in a model that has an overall greater understanding from
a dataset providing a limited number of examples from which to learn
and generalize. In the dense blocks, to prevent redundant learning,
each layer takes what has been previously learned as input. To
increase spatial understanding, dilation is added to the convolutions
to further increase receptive fields at reduced resolutions and
understand where lesions are relative to other lesions during
segmentation. The decoder portion employs a series of up-convolution
and concatenation with extracted high-resolution features from the
encoder in order to gain better localization. The dilated dense UNet
was designed to allow lower level features to be extracted from 2D
input images and passed to higher levels via dense connectivity to
achieve more robust image segmentation.

\begin{table}
\centering
\begin{tabular}{lcccr}
\toprule
Organ & Modality & Model: Dilated Dense UNet \\
\midrule
Brain   & MR      & 0.6093     \\
Lung    & CT       & 0.6978     \\
\bottomrule
\end{tabular}
\caption[Dice score for the dilated dense UNet model on the brain and
lung datasets]{Dice score for the dilated dense UNet model on the
brain and lung datasets.}
\label{table:DenseDilatedUnet}
\end{table}


\subsection{Implementation}

\paragraph{Upconvolution:}
As opposed to the bilinear interpolation proposed by
\citet{ronneberger2015u}, we propose a different method of upsampling
on the decoder side---Transpose convolutions are used to upsample. The
reason is that this method has parameters that can be learned while
training, as opposed to a fixed method of interpolating. This allows
for an optimal upsampling policy.

\paragraph{Dilation:}
Dilated convolutions utilize sparse convolution kernels with large
receptive fields, spatial understanding, and the advantage of few
training parameters. Dilation is added to the bottom of the
architecture, as shown in Figure~\ref{fig:dilateddensunet}.




The last dense block at the end of the contracting path is fed into 4
convolutional layers with dilation 2, 4, 8, and 16. Once the blocks go
through the dilated convolutions, they are concatenated to gain wider
spatial perspective at the end of the contracting path of the dilated
dense UNet. This is only effective and necessary at the end of the
contracting path because this area samples at the lowest resolution
and can lose track of spatial understanding. Therefore expanding the
spatial context with dilation before continuing on the expanding path
is an efficient and effective method to obtain better results.

\subsection{Results and Analysis}

Table~\ref{table:DenseDilatedUnet} shows the results of the dilated
dense UNet model. We observed that the dilated dense UNet model
performed the best compared to the baseline UNet and the Dense UNet.

The dilated dense UNet served as the backbone for a full segmentation pipeline proposed in our CVPR 2019 submission. The
dilated dense UNet produces an initial segmentation map. This
map is fed into two functions to generate two probability feature maps
$\lambda_{1}$ and $\lambda_{2}$, which the ACL employs to produce a
detailed boundary, thus achieving more accurate segmentation results.


\begin{figure}
\centering
\begin{tikzpicture}
\node (A) {\includegraphics[width=\linewidth]{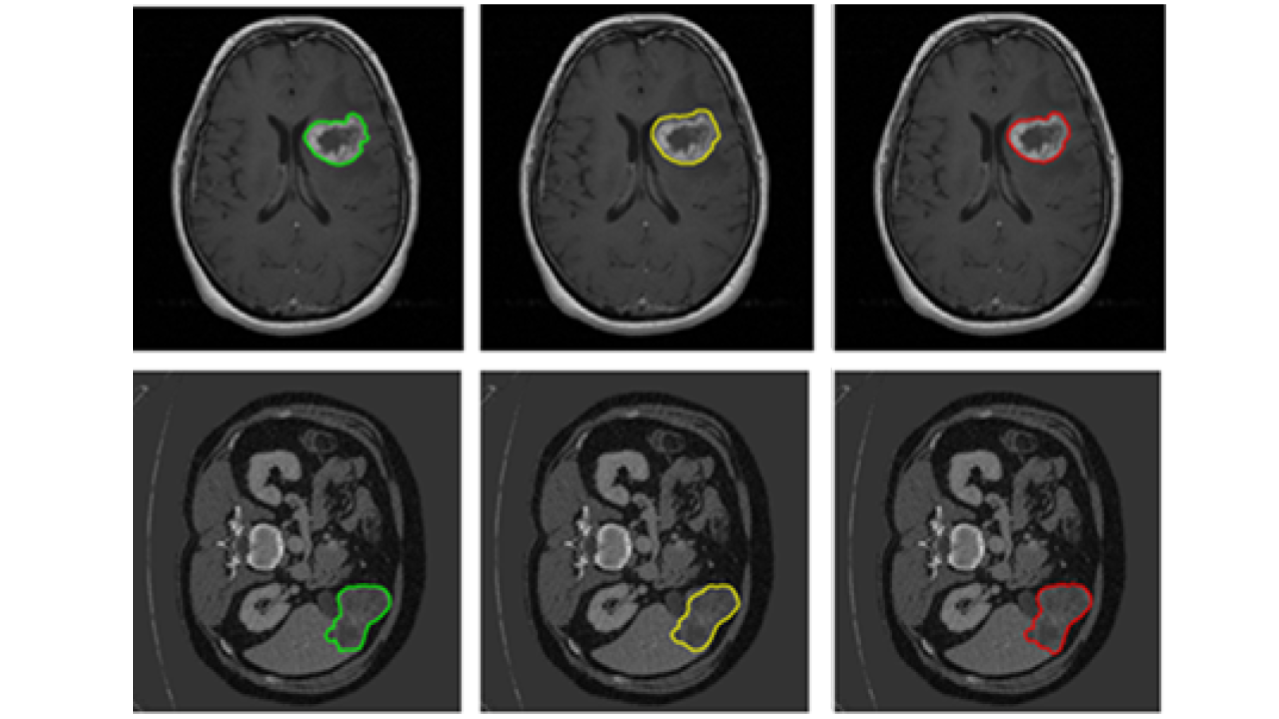}};
\path (A.south west) -- (A.south east) node[pos=.24, below] {(a)} node[pos=.51, below] {(b)} node[pos=.77, below] {(c)};
\end{tikzpicture}
\caption[Examples of Final Segmentation Outputs.]{Examples of final segmentation outputs. (a) Ground Truth (b) Dilated Dense UNet output (c) DLAC output}
  \label{fig:2d_final}
\end{figure}

Figure~\ref{fig:2d_final} shows some examples of the final
segmentations on the Stanford dataset produced by the dilated dense
UNet and dilated dense UNet + ACL models, in comparison to the ground
truth. As a backbone, the dilated dense UNet does an exceptional job
of localizing and determining an initial boundary of the lesions. The
ACL supports the model further to refine the boundaries.
Figure~\ref{fig:2d_final}b validates that the dense dilated UNet can
also be used as an effective, automated, and accurate stand-alone
architecture for lesion segmentation, as well. We also note that this same architecture was successful for two different modalities, namely MR and CT. Thus, we show that our
custom dilated dense UNet is an effective backbone for the
task of lesion segmentation for brain and lung.

\section{Loss Function and Evaluation Metrics}

As the loss function, we utilize the Dice coefficient defined in Equation
\ref{eq:DiceLoss}. It performs exceptionally on problems for which
there is heavy class imbalance in the training dataset. This is indeed
the case for the task of lesion segmentation, as can be seen in
Figure~\ref{fig:2D_seg_pics}. In the case of brain and lung
segmentation, because lesions are very small, there is a clear ``class
imbalance'' between the number of pixels in the background and
foreground of the image.

\section{Effects of Using Pretrained Models}

An issue with segmentation involving medical images is the lack of
large aggregate datasets. The proposed models required 2D images
labeled with binary masks indicating the location of lesions. The
Stanford dataset is rather small to see the full potential of this
model. When the model was initially trained on this small dataset,
nice learning trends were not observed. The model overfit rather
quickly. We hypothesized that this was because there were not enough
examples to properly generalize the model. Therefore, we ran
experiments with and without the use of pretrained models. The results
are reported in Table~\ref{table:pretrain}.

\begin{table}
  \centering
  \begin{tabular}{lcccccccccr}
  \toprule
  Organ        & UNet       & DUNet     & Dilated DUNet    & UNet (P)    & DUNet (P)    & Dilated DUNet (P) \\
  \midrule
    Brain        &    0.5231  &    0.5839 &    0.6093         & 0.5873      & 0.6105       & 0.7541     \\
    Lung         &    0.6646  &    0.6723 &    0.6978         & 0.7137      & 0.7028       & 0.8231      \\
  \bottomrule
  \end{tabular}
  \caption[Model Evaluations]{Model evaluations. Dice values for each
  model are presented; (P) indicates that the model utilized
  pretrained weights. DUNet denotes the dense UNet.}
  \label{table:pretrain}
\end{table}

The segmentation of lesions is a particularly difficult task because
priors cannot be utilized for predicting the shapes of lesions, since
each lesion is unique. The results in Table~\ref{table:pretrain}
indicate that using a pretrained model is clearly an effective
strategy to help the model learn with such a small dataset.
Pretraining can be thought of as a kind of ``prior'' added to the
model to assist in the learning process when training on the dataset
of interest. The pretraining allowed the 2D dilated dense UNet model
to segment lesions with a Dice score of 82\% for lung and 75\% for
brain images.

\chapter{3D Medical Image Segmentation}

In recent years, the rise of deep learning models for 2D medical image
segmentation has been very prevalent and successful. The same trend is
starting to be observed for 3D medical data. After discussing the
implementations of deep learning models for 2D medical image
segmentation, this chapter transitions to exploring and validating the
considerations to take into account when developing deep learning
models for 3D medical image segmentation. There are many advantages to
3D medical image datasets. 3D datasets offer spatial coherence, which
is quite beneficial for segmentation tasks. Although the availability
of 3D data is limited, it provides important information that can help
a deep model learn more accurate segmentation parameters. However,
although 3D data provides rich information unavailable in 2D data, 3D
data poses many challenges and transitioning to 3D is far from
trivial. This chapter presents important considerations for
preprocessing datasets and implementing an efficient 3D medical image
segmentation model. The effectiveness of these considerations are
evaluated by testing them on an abdominal lymph node dataset \citep{seff2015leveraging}. A 3D
VNet model is then trained using this dataset.

\section{Background and Dataset}

Lesion and organ segmentation of 3D CT scans is a challenging task
because of the significant anatomical shape and size variations
between different patients. Much like in the 2D case, 3D medical
images suffer from low contrast from surrounding tissue, making
segmentation difficult.

An interesting application for 3D medical imaging is the segmentation
of CT scans of the human abdomen to detect swollen lymph nodes. Lymph
nodes are small structures within the human body that work to destroy
harmful substances. They contain immune cells that can help fight
infection by attacking and destroying microbes that are carried in
through the lymph fluid. Therefore, the lymphatic system is essential
to the healthy operation of the body. The presence of enlarged lymph
nodes is a signal to the onset or progression of an infection or
malignancy. Therefore accurate lymph node segmentation is critical to
detect life threatening diseases at an early stage and support further
treatment options.

\begin{figure}
\centering
\begin{tikzpicture}
\node (A) {\includegraphics[width=\linewidth]{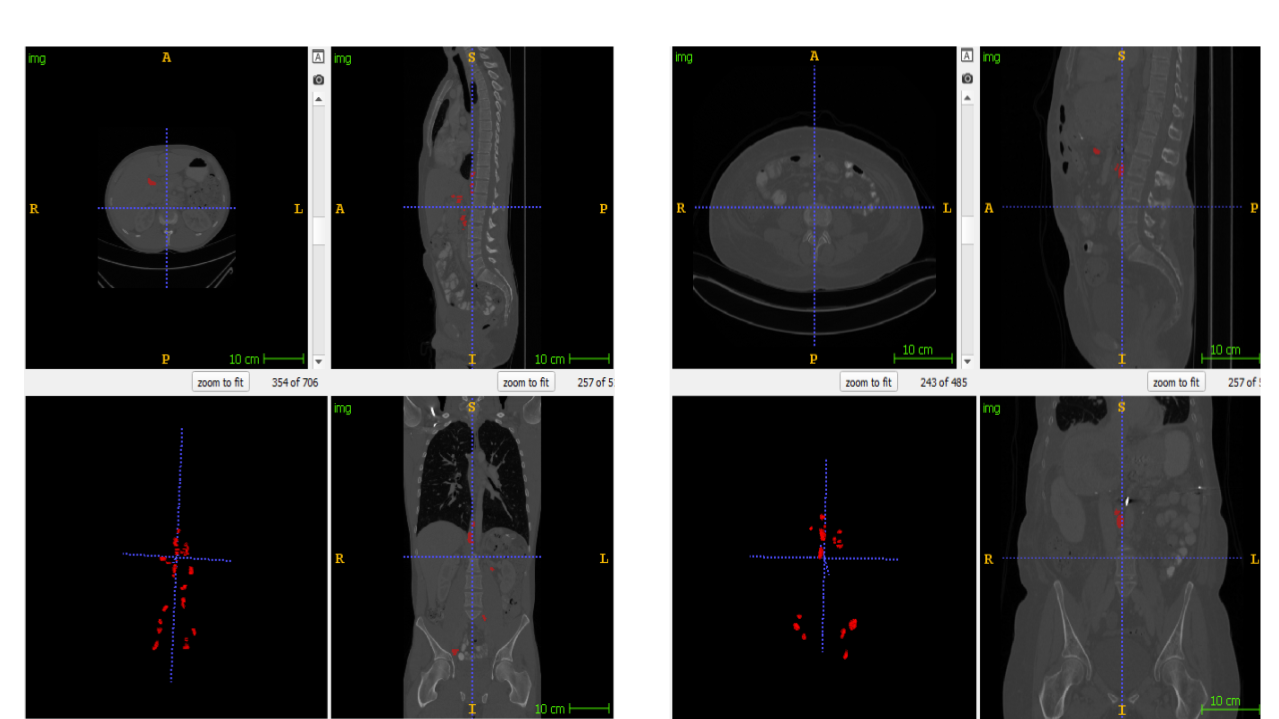}};
\path (A.south west) -- (A.south east) node[pos=.15, below] {(1)} node[pos=.37, below] {(2)} node[pos=.65, below] {(3)} node[pos=.87, below] {(4)};
\node (B) {\includegraphics[width=\linewidth]{Images/3d_lymph.png}};
\path (A.north west) -- (A.south west) node[pos=.26, below] {(a)} node[pos=.76, below] {(b)};
\end{tikzpicture}
\caption[Examples of Lymph Node Segmentation.]{Examples of the top (a1, a3), side (a2, a4) and front view (b2, b4) CT scans. The ground truth masks are depicted in red (b1, b3). The low contrast in the CT scans make distinguishing the lymph node from the surrounding tissue a difficult task, even for the human eye.}
  \label{fig:3d_lymph}
\end{figure}

The task of detecting and segmenting swollen lymph nodes in CT scans
comes with a number of challenges. One of the biggest challenges in
segmenting CT scans of lymph nodes in the abdomen is that the
abdominal region exhibits exceptionally poor intensity and texture
contrast among neighboring lymph nodes as well as the surrounding
tissues, as shown in Figure~\ref{fig:3d_lymph}. Another difficulty is
that low image contrast makes boundary detection between lymph nodes
extremely ambiguous and challenging \citep{nogues2016automatic}.




\section{3D Data Preprocessing Techniques}

There are several key points to consider when developing a 3D
segmentation architecture, aside from how to fit large 3D images into
memory. The first is the preprocessing techniques to be used on the
dataset of interest, including image orientation and normalization
methods. Another important consideration that makes a substantial
difference in performance for 3D segmentation is the loss function
that is utilized for training. Thus, the key considerations are as
follows:
\begin{enumerate}
\item \textbf{Consistent Orientation Across Images:} Orientation is
important in the 3D space. Orienting all images in the same way while
training reduces training time because it does not force the model to
learn all orientations for segmentation.
\item \textbf{Normalization:} This is a key step and there are
differing techniques for different modalities.
\item \textbf{Generating Segmentation Patches:} This is an important
strategy to use if there are memory constraints, which is often the
case for 3D medical images. This strategy also aids in the issue of
class imbalance in datasets.
\item \textbf{Loss Functions and Evaluation Metrics:} Loss functions
are important considerations for 3D segmentation and can also aid in
the issue of class imbalance in datasets.
\end{enumerate}

\begin{figure}
  \includegraphics[width=\linewidth]{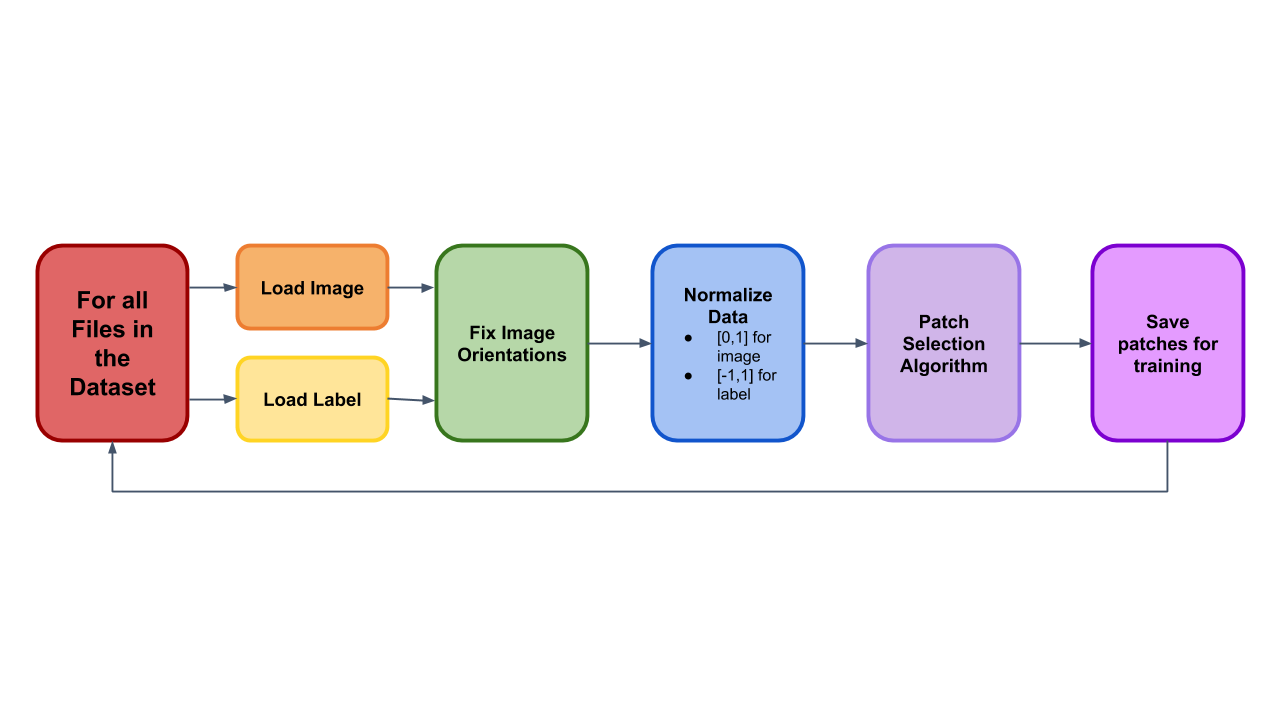}
  \caption{Preprocessing pipeline for 3D dataset.}
  \label{fig:preprocess}
\end{figure}

To alleviate the issues listed above and properly preprocess the 3D
data, the lymph node dataset was fed through the pipeline shown in
Figure~\ref{fig:preprocess}. The next sections describe the
preprocessing steps in more detail.

\subsection{Consistent Orientation}

In 3D images, the idea of orientation becomes an issue of interest.
Consistent orientation among all images is important to speed up
training time. If there are multiple orientations, the network is
forced to learn all orientations, and in order to generalize, more data
is required, which is not easily available for 3D medical images.
Nibable, an open source python library to read in medical images, was
utilized to determine 3D image orientation and reorient images if
necessary. It was found that the orientation itself did not make a
difference on training so long as the orientation was consistent
throughout the dataset.

\subsection{Normalization}

Normalization is a key step in the preprocessing pipeline for any deep
learning task. Normalization is also very important for medical images
and there are a variety of methods for doing this. The aim of
normalization is to remove heavy variation in data that does not
contribute to the prediction process and instead accentuate the
features and differences that are of most importance. The following
methods may be used specifically for medical image segmentation
\citep{dltk_overview}:
\begin{enumerate}
  \item \textbf{Voxel Intensity Normalization:} This method is very
  dependent on the imaging modality. For images such as weighted brain
  MR images, a zero-mean unit variance normalization is the standard
  procedure. This is done because the contrast in the image is usually
  set by an expert taking the MR images and thus there is high
  variation in intensity across image sets. This variation may be
  considered noise. To standardize intensity across multiple intensity
  settings, we use a zero mean normalization. In contrast, CT imaging
  measures a physical quantity such as radio-density in CT imaging,
  where the intensities are comparable across different scanners.
  Therefore, for this modality, the standard normalization methodology
  is clipping or rescaling to a range such as $[0,1]$ or $[-1,1]$.
  
  \item \textbf{Spatial Normalization:} Normalizing for image
  orientation avoids the need for the model to learn all possible
  orientations of input images. This reduces both the need for a large
  amount of training data and training time. Since, as mentioned
  previously, properly labeled medical data is often meagre in
  quantity, this is a very effective technique of normalization.
\end{enumerate}

Since the the modality of the abdominal lymph node data was 3D CT
scans, both methods of normalization were used. Reorientation was also
done on all images. Voxel intensity normalization was performed by
rescaling all voxels in the images to a range of $[-1,1]$ and the
voxels in the labels to $[0,1]$.

\subsection{Generating Segmentation Patches}

Patch-based segmentation is used to tackle the issue of limited
memory. A single 3D image in the lymph node dataset was $512 \times
512 \times 775$. With the memory constraints, using these raw images
would only allow for a batch size of 1, which would result in an
incredibly long training time, and very little flexibility to extend
the deep learning architecture. Therefore patch-based segmentation was
utilized.

The basic idea of patch generation is to take in an input image,
determine the regions of greatest interest in the image, and return a
smaller portion of the image, focusing on these regions of importance.
Doing this is important for many reasons. The first reason is that
having smaller image sizes allows for larger batch sizes and faster
training. The second reason is that the data is naturally augmented
with this process. The third reason is that class imbalance issues can
be avoided. Class imbalance occurs when a sample image has more voxels
belonging to one class over another. Figure~\ref{fig:3d_lymph} shows
this very clearly. In the example, the majority of the voxels belong
to the background (black) class and a small subset belong to the
foreground (red) class. Feeding such images directly to the model will
cause the the model to skew its learning to favor the background
voxles. Therefore, intelligent selection of patches is key to
training. For this work, we generated patches of $128 \times 128
\times 128$. We utilized the Deep Learning Toolkit to generate our
class-balanced segmentation patches.


Unlike during training time, in which the patches are generated
randomly around areas of interest, during test time, the images are
systematically broken into $128 \times 128 \times 128$ chunks.
Prediction is done on each chunk and the chunks are joined together
for the final prediction. The patching technique alone creates some
discrepancies at the boarders of patches, so smoothing is performed to
obtain a more seamless prediction.

\subsection{Loss Functions and Evaluation Metrics}

The choice of loss functions can significantly affect the performance
of 3D segmentation architectures. In medical images, it is very common
that the region of interest, for example, a lesion, covers a very
small portion of the scan. Therefore, during the learning process, the
network often gets stuck in local minima due to this class imbalance
of background and foreground pictures. This causes poor results during
training and testing since the prediction of both classes is weighed
equally. This causes the model to be biased toward the background
voxels and the foreground is either completely missed or partially
detected. The following losses attempt to address these issues.


\paragraph{Jaccard Coefficient and Loss}

The Jaccard coefficient is measure of similarity between a ground truth mask and a prediction mask. Lets assume that all voxels in the ground truth mask are in set $G$. All voxels in the prediction mask are in set $P$. The Jaccard coefficient calculates the ratio of 3 mutually exclusive partitions: $|P \cap G|$, $|P \setminus G|$, $|G \setminus P|$. The equation for the Jaccard coefficient is
\begin{equation}
\begin{split}
\label{eq: jaccard_c}
J = \frac{|P \cap G|}{|P \cap G| + |P \setminus G| + |G \setminus P|},
\end{split}
\end{equation}

and the full loss function is defined as $1-J$, where the $|P \setminus G|$
notation refers to the set of elements in $P$ but not in $G$. The Jaccard coefficient calculates the overlap ratio between a ground truth mask and a predicted mask by giving equal weightage to  $|P \cap G|$, $|P \setminus G|$, $|G \setminus P|$. If the data of interest does not suffer too much from class imbalance, the Jaccard coefficient and loss is a good metric to measure the success of a model. If there is class imbalance, this method may not produce as promising results and a model could become biased towards the dominating class. 

\paragraph{Dice Coefficient and Loss:}
The Dice coefficient is a metric that compares the similarity between
a ground truth mask and a prediction mask. It is similar to the
Jaccard coefficient with one subtle but significant difference---in
the way class imbalance is handled. Dice Loss was used in Chapter~4,
as it is a common loss function for both 2D and 3D medical image
segmentation. The Dice coefficient, also known as the Sorensen-Dice
coefficient, is defined as
\begin{equation}
\label{eq: Dice_c}
D = \frac{2 \times |P \cap G|}{2 \times |P \cap G| + |P \setminus G| +
|G \setminus P|}
\end{equation}
and the Dice Loss is defined as $1 - D$. The Dice
coefficient is a measure of similarity between two sets. The Dice
coefficient addresses the issue of class imbalance by calculating the
overlap between a prediction and a ground truth mask for a region of
interest. The Dice coefficient is unique because unlike the Jaccard
metric, the Dice metric gives more weight to voxels that belong to
both sets (a.k.a. the overlapping region of both sets) when
calculating the overlap ratios, hence the factor of 2 that is
multiplied to the union sum in the numerator. For this reason, the
Dice metric and loss function is often utilized for medical image
segmentation since correct overlapping regions are hard to find due to
irregular lesion shapes.

\paragraph{Tversky Coefficient and Loss:}

The Tversky coefficient is yet another similarity metric related to
both the Jaccard and Dice coefficients with a different method of
handling class imbalance. The Tversky Coefficient is defined as
\begin{equation}
\label{eq: tversky_c}
T = \frac{|P \cap G|}{|P \cap G| + \alpha |P \setminus G| + \beta |G
\setminus P|}
\end{equation}
and the Tversky Loss is defined as $1 - T$, where $\alpha$ and $\beta$
control the magnitude of penalties of false positive versus false
negative errors in the pixel classification for segmentation. The
ranges of the $\alpha$ and $\beta$ hyperparameters, which can be tuned
to produce optimal loss results on the dataset of interest, must be
such that $\alpha + \beta = 1$. The Tversky metric provides a spectrum
of methods to normalize the size of a two-way set intersection. The
Tversky index can be thought of as a generalized Dice index, since the
Dice coefficient is the Tversky coefficient with $\alpha = \beta =
0.5$.

\section{3D VNet Implementation and Data Processing Pipeline}

Our base implementation of the VNet architecture was inspired by the
work of \citet{jackyko1991_vnet_tensorflow}. The main portions of this
architecture involved a data-loader, the VNet model, and an evaluation
pipeline to see how well the model can predict on new images. However,
the pipeline had to be heavily modified. In medical imaging, data input-output
is often one of the most challenging steps. This is especially
challenging for the case of 3D images, since special care must be
taken to load and preprocess the data. The initial data-loader and
training pipeline involved using the SimpleITK package to read in,
normalize, and randomly select patches on which to train the model.
The disadvantage to using SimpleITK is that this library does not
utilize GPU-accelerated operations, making the data-loader slow and
inefficient. SimpleITK was also utilized for the evaluation pipeline.
Since SimpleITK is not GPU-compatible, memory constraints were an
initial problem. Therefore, the data-loader and evaluation pipelines
were then changed to utilize the Nibabel and Deep Learning Toolkit
(DLTK) packages. The Nibabel package was used to read in a image,
normalize the image, and orient all images in the same way. The DLTK package
was used to select patches on which to train the model. DLTK utilizes
an algorithm that determines the regions of interest in each image and
creates patches around this region to promote class-balanced samples.
This improved end-to-end pipeline was utilized to train the VNet model
to predict swollen lymph nodes in 3D abdominal CT scans.

\subsection{Results and Analysis}


\begin{figure}
\centering
\begin{tikzpicture}
\node (A) {\includegraphics[width=\linewidth]{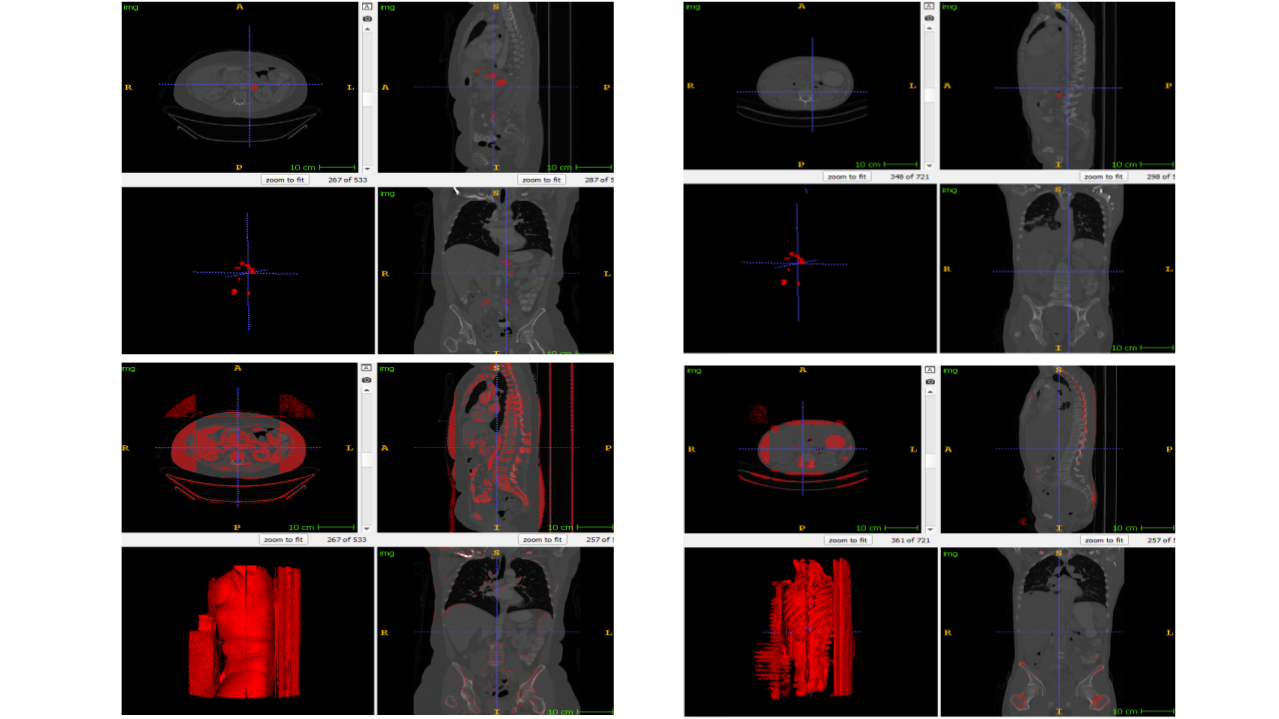}};
\path (A.south west) -- (A.south east) node[pos=.29, below] {(1)} node[pos=.73, below] {(2)};
\node (B) {\includegraphics[width=\linewidth]{Images/lymph_final.png}};
\path (B.north west) -- (B.south west) node[pos=.25, below] {(a)} node[pos=.73, below] {(b)};
\end{tikzpicture}
\caption[3D Segmentation Results with VNet architecture and Dice Loss.]{(a)Examples of segmentation overlaid with ground truth masks. (b) Examples of segmentation overlaid with predicted masks. In each set of 4 images, the top left is a top view, top right is a side view, bottom left is the segmentation mask, bottom right is a front view. (1) shows results from a few hours of training. (2) show results from a few days of training.}
  \label{fig:lymph_final}
\end{figure}



Figure~\ref{fig:lymph_final} shows our preliminary results of this
model on the lymph node dataset. The 3D VNet model was trained using a
Dice loss function. Figure~\ref{fig:lymph_final}a depicts the CT scans
overlaid with the ground truth masks. Within each group of four images, the lower left depicts the 3D
isolated lymph nodes. Figure~\ref{fig:lymph_final}b depicts the CT
scans overlaid with the predicted masks from this VNet architecture.
Again, in the groups of four, the lower left image shows the isolated 3D mask prediction. For
Figure~\ref{fig:lymph_final}a, it is evident how difficult this
problem is, due to small size of the lymph nodes compared to the rest
of the body. These preliminary results were obtained by training 1. for few hours (Figure \ref{fig:lymph_final} (1)) and 2. for a few days (Figure \ref{fig:lymph_final} (2)). An interesting observation is
that even within a few hours, the VNet started to distinguish
boundaries in the image, which explains why the bottom left image in
Figure~\ref{fig:lymph_final} (b1) look like the outline of the patient in
3D. After training the model for a few days, it was observed that even
more details are learned by the VNet, such as the ribs
(Figure~\ref{fig:lymph_final} (b2)). This trend suggests that if the
model was trained long enough, or was trained using more computational resources, it would start to pick up the nuances
of the lymph nodes. However, this also suggests that there are many
improvements that could be made to this model to enable better
performance, such as adding dense connectivity to the VNet, similar to
the 2D version in Chapter~4, different loss functions, or different
styles of encoder-decoder architectures. Due to time constraints, we
were unable to be perform exhaustive evaluation metrics and trials on
the effects of different loss functions for this dataset. However,
this work has set the foundation for many future endeavors and
extensions of our work.

\chapter{Conclusion}

Image segmentation is a topic of great interest in computer vision and
it has been utilized for a variety of tasks, such as object
localization and recognition, boundary detection, autonomous driving,
and medical imaging. Prior successful segmentation methods relied on
some level of manual user initialization. This thesis proposed
effective architectures to fully automate the process of localizing
and delineating boundaries of objects in images. In particular, we
developed and implemented deep learning frameworks for two
segmentation tasks, namely salient object detection and 2D medical
image segmentation. We also explored and evaluated the considerations
that should be taken into account when moving from 2D medical image
segmentation to 3D, thereby setting the foundation for future
extensions of our 2D models to 3D.

First, we addressed the problem of salient object detection. Several
encoder-decoder architectures were described before arriving at a
novel dilated dense encoder-decoder network with dilated spatial
pyramid pooling. Serving as the backbone CNN architecture in
conjunction with an active contour layer, our full pipeline
outperformed on one of the three popular datasets used for salient
object detection, and was very competitive on the other two datasets.
The segmentation results of the dilated dense encoder-decoder itself
were also impressive, indicating that this architecture can serve as a
robust stand-alone segmentation architecture as well.

Next, we addressed an important and relevant use case of segmentation,
namely 2D medical image segmentation. We developed a novel dilated
dense UNet architecture to combat the many challenges faced with 2D
lesion segmentation, including low image contrast, lack of large
labeled datasets, and the lack of robust pretrained models. A custom
dataset from a collaboration with Stanford University was utilized to
evaluate our model's performance. Compared to the baseline, our
dilated dense UNet model demonstrated significantly better
performance, confirming the effectiveness of our architecture. Our
architecture was used as the backbone CNN for accurate initial lesion
localization and boundary delineation in a full pipeline, which also combined active contours. The segmentation results of the
stand-alone dilated dense UNet also showed promising results for 2D
lesion segmentation, suggesting that it can perform well on its own.

\section{Future Work}

Our discussion in Chapter~5 set the foundation for developing a robust
3D model for medical image segmentation, which is important since
modern medical imaging systems often acquire 3D images. There are many
possible improvements to our model, including exploring additions to
the basic VNet architecture, how different loss functions affect
training, or experimenting with the effectiveness of using greater
computational resources, if available.

\subsection{Additional Loss Functions}

For the VNet model, a Dice loss function was utilized. This was chosen
because it is the most popular metric for the case of medical image
segmentation. However, one should experiment with other loss functions
to determine which is most effective at handling class imbalance in
datasets. For the case of lymph node segmentation, there was heavy
class-imbalance, as observed in Figure~\ref{fig:lymph_final}a.
Therefore, a useful loss function to explore could be the Tversky loss
function. Weighted Cross Entropy is yet another effective loss
function utilized in more general segmentation tasks. Multi-layered loss functions are another interesting option to explore.


\subsection{Additions to the VNet Architecture}

The first suggested modification, in conjunction to the previous
chapters of this work, is to experiment with adding dense blocks to
the VNet architecture. As seen in the 2D case, it is reasonable to
hypothesize that adding dense connectivity will increase the learning
capabilities of the model because each layer will directly take as
input all previously learned feature maps. For challenging
segmentation tasks, such as swollen lymph node segmentation, we
observed that the learning process is rather slow. Dense connectivity
could also help speed up the learning process due to the easier flow
of gradients during backpropagation. Another suggested modification to
the VNet architecture is to add dilated convolutions to gain spatial
awareness on the encoder side of the VNet. Dilated convolutions can be
useful in the 3D case because they are computationally efficient with
the advantage of an expanded receptive field, which is critical in the
3D case.

\subsection{Additional Computational Resources}

All the work presented in this thesis utilized a single Nvidia 1080 Ti
GPU. This was a powerful resource effectively training the 2D models
developed in Chapters~3 and 4 on $256 \times 256$ and $512 \times 512$ images. For the 3D datasets, the image sizes varied from $512 \times 512 \times 48$ to
$512 \times 512 \times 775$. These images were much too large to fit
more than one into memory at a time. Therefore, a patch based method
was adopted to train the segmentation model. Although our preliminary
results suggested the potential effectiveness of the patch based
approach to training, given greater resources permitting the
accommodation of an entire image for training, it seems reasonable to
expect that even a basic VNet model would be much more effective and
robust. Regardless of the computational resource limitation, however,
our work demonstrates a series of highly effective 2D models and sets
the stage for extending them to the purposes of 3D medical image
analysis.

In the recent past, a number of tasks dependent on image segmentation have come to light, such as medical imaging, autonomous driving, or scene understanding. This research work hoped to gain a deeper understanding of image segmentation and contribute to this topic, which has become critical to understanding and assisting the world around us.

\bibliographystyle{apa}
\bibliography{thesis}

\end {document}